\documentclass[a4paper,11pt]{article}

\usepackage{jheppub} 
\usepackage{subfigure}
\newcommand{\tophat}[2]{\mathcal{W}_{#1}^{#2}}

\preprint{KUNS-3072}

\title{\boldmath 
Physics-informed neural network solves minimal surfaces in curved spacetime
}

\author[a]{Koji Hashimoto,}
\author[a]{Koichi Kyo,}
\author[b]{Masaki Murata,}
\author[b]{Gakuto Ogiwara,}
\author[a]{Norihiro Tanahashi}

\affiliation[a]{Department of Physics, Kyoto University, Kyoto 606-8502, Japan}
\affiliation[b]{Department of Information Systems, Saitama Institute of Technology, Saitama
369-0293, Japan}

\emailAdd{koji@scphys.kyoto-u.ac.jp}
\emailAdd{kyo.koichi@gauge.scphys.kyoto-u.ac.jp}
\emailAdd{m.murata@sit.ac.jp}
\emailAdd{i5001hrr@sit.ac.jp}
\emailAdd{tanahashi@gauge.scphys.kyoto-u.ac.jp}

\abstract{
We develop a flexible framework based on physics-informed neural networks (PINNs) for solving boundary value problems involving minimal surfaces in curved spacetimes, with a particular emphasis on singularities and moving boundaries. By encoding the underlying physical laws into the loss function and designing network architectures that incorporate the singular behavior and dynamic boundaries, our approach enables robust and accurate solutions to both ordinary and partial differential equations with complex boundary conditions.
We demonstrate the versatility of this framework through applications to minimal surface problems in anti-de Sitter (AdS) spacetime,
including examples relevant to the AdS/CFT correspondence (e.g.~Wilson loops and gluon scattering amplitudes)
popularly used in the context of string theory in theoretical physics.
Our methods efficiently handle singularities at boundaries, and also support both ``soft" (loss-based) and ``hard" (formulation-based) imposition of boundary conditions, including cases where the position of a boundary is promoted to a trainable parameter.
The techniques developed here are not limited to high-energy theoretical physics but are broadly applicable to boundary value problems encountered in mathematics, engineering, and the natural sciences, wherever singularities and moving boundaries play a critical role.}

\begin{document} 

\maketitle
\flushbottom

\section{Introduction}
\label{sec:intro}

Minimal surfaces, or extremal surfaces more generally, play a central role in a wide variety of scientific fields. In mathematics, minimal surfaces are a classical subject in differential geometry and the calculus of variations. They also describe minimal-energy configurations of membranes, which have applications in various fields, including engineering and the life sciences.

Beyond these traditional fields, minimal surfaces have been attracting interest in various research areas in theoretical physics. Minimal surfaces are one of the most fundamental geometric objects, and hence they appear in gravitation theory in multiple ways. For example, the black hole horizons can be regarded as minimal surfaces, and the gravitational dynamics of membrane-like objects are also active research targets. Such membrane-like objects typically also appear in high-energy physics and string theory. Physical quantities such as the entanglement entropy, the Wilson loops, and computational complexities in quantum field theory are all related to minimal surfaces in curved spacetime via the AdS/CFT correspondence \cite{Maldacena:1997re}, which connects quantum field theory with theories of gravity and geometry.

The minimal surface configurations are determined by solving the Euler-Lagrange equations, which follow from the variational principle, under given boundary conditions. The practical evaluation of minimal surfaces in curved geometries or under complicated boundary conditions presents significant analytical and numerical challenges, especially when the governing equations take the form of nonlinear partial differential equations (PDEs) with multiple, possibly moving, boundaries.

Recent advancements in machine learning, and in particular the development of physics-informed neural networks (PINNs), have provided powerful new tools for solving such complex PDEs~\cite{RAISSI2019686,
NASCIMENTO2020103996}. PINNs incorporate the underlying physical laws as part of the loss function, allowing the solution to satisfy both the governing equations and diverse constraints such as boundary conditions. This approach is well-suited for constructing minimal surfaces with complex boundary conditions, as it flexibly handles intricate geometries, nonlinearities, and non-standard boundary conditions.

In recent studies PINNs have been applied to minimal-surface problems in flat or Euclidean domains, including high-dimensional settings and tensile membrane form-finding, by embedding the Euler–Lagrange equations directly into the loss function~\cite{zhou2023approximatinghighdimensionalminimalsurfaces, Kabasi2023PINN}. General-purpose PINN libraries have facilitated these developments~\cite{peng2021idrlnetphysicsinformedneuralnetwork}, and related machine-learning approaches have also been explored for minimal-surface-like structures such as triply periodic minimal surfaces (TPMS) \cite{Mishra2025TPMS}. However, to the best of our knowledge, there has been no prior PINN-based study of minimal surfaces embedded in curved spacetimes. This work fills that gap by formulating and solving minimal-surface boundary value problems in anti-de Sitter (AdS) spacetimes, where the asymptotic boundary induces a singularity in the Euler–Lagrange equation and additional domain walls introduce moving or Neumann-type boundary conditions.

In this work, we demonstrate the versatility and effectiveness of PINNs for solving minimal surface problems in curved spacetimes. As a showcase of its versatility, we apply this technique to minimal surfaces in a curved spacetime called the anti-de Sitter (AdS) spacetime. This AdS spacetime has an outer asymptotic boundary, which appears as a singularity in the Euler-Lagrange equation governing the minimal surface. We also consider the case where the spacetime has an additional wall at a finite distance from the asymptotic boundary. The minimal surface is required to intersect this wall perpendicularly, and from the technical point of view, the problem becomes a boundary value problem with a moving boundary when formulated in spherical coordinates. We demonstrate that the numerical method based on PINN can efficiently handle singularities and moving boundaries, making it a valuable alternative to conventional numerical techniques.

The minimal surface problems in AdS spacetimes with the moving/singular boundaries are adopted in this paper because the problems are popular in high energy physics through the AdS/CFT correspondence \cite{Maldacena:1997re}. In fact, the two-dimensional minimal surface appears in the evaluation of physical observables such as Wilson loops \cite{Maldacena:1998im, Rey:1998bq} and gluon scattering amplitudes \cite{Alday:2007hr}. In certain phases, the corresponding minimal surfaces need to end on other minimal surfaces called D-branes, thus the singular/moving boundaries are additionally introduced. Therefore, the problems described and solved in this paper are for physical problems and serve as a benchmark for realistic situations in physics research. In fact, in our companion paper~\cite{Hashimoto2025instanton} we study the instanton corrections to the gluon scattering amplitudes in the AdS/CFT correspondence, which boils down to a particular minimal surface problem studied in this paper. 

This paper is organized as follows. In section~\ref{sec:WL}, we present a detailed analysis of minimal curves in AdS spacetime. We introduce the action, derive the equations of motion, and study the relevant boundary conditions, followed by both exact and numerical solutions using the PINN framework.
For the problem with the Neumann boundary, there are various methods to implement the boundary conditions.
We present and compare these various methods in terms of, for example, code structure and calculation cost.
Section~\ref{sec:minimal-surface} extends the study to two-dimensional minimal surfaces in curved space, describing the setup and numerical methodology, and illustrating the PINN approach for both standard and Neumann-type boundary conditions.
%
Section~\ref{sec:light-like_loop} applies the framework to minimal surfaces bounded by a light-like polygonal loop. 
%
This problem is a boundary problem for a set of two nonlinear elliptic PDEs, where the boundary condition contains stronger singularities than those in the previous problems. We report our attempt to overcome such difficulties.
We also examine the case with a Neumann boundary in this setup. 
%
%
Finally, section~\ref{sec:summary} is devoted to a summary of our results and a discussion of possible future directions.
Appendix~\ref{App:ansatz} provides a detailed derivation of the solution ansatz used in section~\ref{sec:light-like_loop}.
In appendix~\ref{app:square-alt}, we discuss an issue on the choice of PINN architecture for the problem in section~\ref{sec:light-like_loop}.

The physical motivation for the setups treated in section \ref{sec:WL} and \ref{sec:minimal-surface} is the holographic Wilson loops in the AdS/CFT correspondence \cite{Maldacena:1998im}, and that in section \ref{sec:light-like_loop} is 
the holographic gluon scattering amplitude \cite{Alday:2007hr}. In particular, the setup in section \ref{sec:light-like_loop} with the Neumann boundary 
corresponds to taking into account the effect of an instanton in the calculation of the holographic gluon scattering amplitude, providing a numerical derivation of the minimal surface studied in our companion paper~\cite{Hashimoto2025instanton}. The latter serves as a concrete showcase of PINN as a flexible AI solver of physics.


\section{Minimal curve in curved space}
\label{sec:WL}

\subsection{Action and equation of motion} 
\label{sec:WL-setting}

As the simplest problem to construct a minimal surface in the curved spacetime, 
we consider a physics problem of a holographic Wilson loop in the AdS/CFT correspondence \cite{Maldacena:1998im}. This problem is essentially reduced to finding the shape of a one-dimensional string with constant tension whose ends are attached to the boundary of the curved spacetime called the anti de-Sitter (AdS) spacetime.

We consider a Wilson loop associated with two static points separated by a distance $L$ in the $x$ direction. Then, the holographic dual of the Wilson loop is given by a Nambu-Goto surface in AdS spacetime, where only its AdS$_3$ part is relevant to the following calculations:
\begin{equation}
ds^2 
= g_{\mu\nu} dx^\mu dx^\nu
= \frac{1}{z^2} \left(
-dt^2 + dx^2 + dz^2
\right) \, .
\label{AdS3}
\end{equation}
This spacetime has a boundary at $z=0$, which is called the AdS boundary, at which the spacetime ends.
In this spacetime, we consider a two-dimensional static Nambu-Goto surface whose target-space coordinates are specified as
\begin{equation}
T = T(\tau)\,,\quad
X = X(\sigma)\,,\quad
Z = Z(\sigma)~.
\end{equation}
Taking the static gauge $T=\tau$, we find the induced metric on the string worldsheet is then given by
\begin{equation}
ds^2
=
g_{\mu\nu}
\partial_a X^\mu \partial_a X^\nu 
d\sigma^a d\sigma^b
= \frac{1}{Z^2} \left[
-d\tau^2 
+ \bigl(
\dot X^2 + \dot Z^2 
\bigr) d\sigma^2
\right]\,,
\end{equation}
where $\dot f := df/d\sigma$.
Then, the Nambu-Goto action is given by
\begin{equation}
S = \int d\tau d\sigma
\sqrt{-\det \bigl( g_{\mu\nu}\partial_a X^\mu \partial_a X^\nu \bigr)}
=
T \int d\sigma \frac{1}{Z^2}
\sqrt{\dot X^2 + \dot Z^2}
=: T \int d\sigma L\,
\label{NGaction_string}
\end{equation}
where $L$ is the proper distance of the Nambu-Goto string on a $t$-constant time slice.
The Euler-Lagrange equation for \eqref{NGaction_string} with respect to $X(\sigma)$ and $Z(\sigma)$ becomes equivalent to each other and is given by 
\begin{equation}
2\dot X^3 - Z \dot Z \ddot X + \dot X \bigl(
2\dot Z^2 + Z \ddot Z
\bigr) = 0\,.
\label{EoM_string}
\end{equation}


The action (\ref{NGaction_string}) has a gauge degree of freedom to rescale $\sigma \to \sigma_\text{new}(\sigma)$.
In this work, we employ the polar coordinates given by
$\sigma = \theta$ and express $(X,Z) = \bigl(R(\theta)\cos\theta, R(\theta)\sin\theta \bigr)$, for which \eqref{EoM_string} reduces to an ODE given by
\begin{equation}
    R'' + R
    - 2 R' \cot\theta\left(
        1 + \frac{R'^2}{R^2}
    \right)
    = 0 \,.
    \label{EoM_string-R}
\end{equation}

\subsection{Boundary conditions}
\label{sec:WL-BC}

We assume that the two ends of the Nambu-Goto string are attached to the AdS boundary ($z=0$) at $x = \pm L/2$.
Due to the reflection symmetry with respect to the $z$ axis, we may limit the coordinate region to $\theta \in [0,\pi/2]$.
Then, the boundary conditions for the string that has two ends at $x = \pm L/2$ are given as follows.
\begin{itemize}
    \item \textit{Dirichlet condition at the AdS boundary}:
    \begin{equation}
        R(\theta=0) = \frac{L}{2}\,.
        \label{Dirichlet-BC_WL}
    \end{equation}
    Under this condition, we can solve the equation of motion~\eqref{EoM_string-R} order by order in $z$ near $z=0$ to construct a series solution given by
    \begin{equation}
        R(\theta) = \frac{L}{2}
        \left(
            1 + \frac12 \theta^2 
        \right)
         + c_3\, \theta^3
         + \mathcal{O}\bigl( \theta^4\bigr)\,,
         \label{Dirichlet-BC_WL2}
    \end{equation}
    where $c_3$ is a constant that cannot be determined only by the Dirichlet boundary condition at $\theta = 0$. This constant $c_3$ is fixed by solving the equation of motion~\eqref{EoM_string-R} under a boundary condition at the other end of the calculation domain.
 
    \item \textit{Neumann condition at $\theta = \pi/2$}:
    
    When we impose the reflection symmetry with respect to the $z$ axis, we need to impose the Neumann boundary condition at $\theta = \pi/2$, which is given by
    \begin{equation}
        R'(\theta=\pi/2)=0\,.
       \label{Neumann-BC_WL}
    \end{equation}
 
\end{itemize}

{\noindent \textit{Neumann boundary at $z=z_0$}: }
In this work, we also consider another problem in which the string has one end at the AdS boundary and the other end is attached to a wall located at $z=z_0$ for some constant $z_0$.
In this case, the variation principle for the action~\eqref{NGaction_string} implies the Neumann condition to be imposed at $z=z_0$, that is,
\begin{equation}
    \frac{d X}{dZ} \biggr|_{z = z_0}
    \propto
    \frac{d}{d\theta} \left( R \cos\theta \right)\biggr|_{R(\theta)\sin\theta = z_0} = 0\,.
    \label{Neumann-BC_WL-z0}
\end{equation}
For convenience, we introduce an angle $\theta = \theta_0$ at which the Neumann condition is imposed, that is,
\begin{equation}
    R(\theta_0) \sin\theta_0 = z_0\,.
    \label{theta0-def_WL-z0}
\end{equation}
Then, the Neumann condition \eqref{Neumann-BC_WL-z0} at $z=z_0$ is expressed as
\begin{equation}
    \frac{d}{d\theta} \left( R \cos\theta \right)\biggr|_{R(\theta)\sin\theta = z_0}
    =
    R'(\theta_0) \cos\theta_0 - R(\theta_0) \sin\theta_0
    = 0 \,.
    \label{Neumann-BC_WL-z0_2}
\end{equation}

\subsection{Exact solution}
\label{sec:WL-exact}

The Nambu-Goto string introduced in section~\ref{sec:WL-setting} admits an exact solution. We summarize its explicit form along with its derivation in this section.

Instead of working on the equation of motion~\eqref{EoM_string-R} in the polar coordinates, we start with the equation of motion~\eqref{EoM_string} in the general gauge and take the following coordinate condition:
\begin{equation}
    \frac{d}{d\sigma} \sqrt{\dot X^2 + \dot Z^2} = 0
    \qquad\Leftrightarrow\qquad
    \dot X\ddot X + \dot Z \ddot Z = 0~.
    \label{gauge_string}
\end{equation}
Under this condition, $\sigma$ becomes proportional to the distance along the string
measured using the flat metric instead of the AdS metric (\ref{AdS3}).


Under the gauge condition \eqref{gauge_string}, the equation of motion \eqref{EoM_string} is simplified as follows:
\begin{equation}
\frac{\delta L }{\delta X}
= - \frac{d}{d\sigma} \frac{\partial}{\partial \dot X}
\left(\frac{1}{Z^2}\sqrt{\dot X^2 + \dot Z^2}\right)
= - \frac{d}{d\sigma}
\frac{\dot X}{Z^2\sqrt{\dot X^2 + \dot Z^2}}
\propto
\frac{d}{d\sigma}
\frac{\dot X}{Z^2}
\propto
Z \ddot X - 2 \dot X \dot Z
=0
\,.
\label{EoM2_string}
\end{equation}

An exact solution of Eqs.~\eqref{gauge_string} and \eqref{EoM2_string} can be derived as follows.
\begin{itemize}
    \item \textit{Standard Nambu-Goto string}:

We construct an exact solution for the Dirichlet boundary condition \eqref{Dirichlet-BC_WL} below.
First, the gauge condition \eqref{gauge_string} implies that we may parameterize $\dot X, \dot Z$ without loss of generality as
\begin{equation}
    \bigl(\dot X , \dot Z\bigr) = \bigl(- A \sin\Theta(\sigma), A \cos \Theta(\sigma)\bigr)\,,
    \label{Thetadef}
\end{equation}
where $A$ is a constant.
Based on the boundary condition \eqref{Dirichlet-BC_WL}, we demand that $\Theta(0) = 0$ and $\Theta(\pi) = \pi$.

To proceed, we rewrite the equation of motion \eqref{EoM2_string} as
\begin{equation}
    Z = \frac{2\dot X \dot Z}{\ddot X}
    \quad \Rightarrow \quad
    \dot Z = \frac{d}{d\sigma}
    \left(
    \frac{2\dot X \dot Z}{\ddot X}
    \right)\,.
\end{equation}
Substituting \eqref{Thetadef}, an equation to fix $\Theta(\sigma)$ is obtained as
\begin{equation}
    2\ddot\Theta\sin\Theta - \dot\Theta^2 \cos\Theta = 0
    \quad\Leftrightarrow\quad
    \frac{d}{d\sigma}
    \left(
    \dot\Theta^{-2}\sin\Theta
    \right) = 0\,.
\end{equation}
This equation can be integrated twice as follows.
\begin{equation}
\dot\Theta = c_1 \sqrt{\sin\Theta}
\quad\Rightarrow\quad
\int \frac{1}{\sqrt{\sin\Theta}} d \Theta = \int c_1 d\sigma = c_1 \sigma + c_0\,.
\end{equation}
The integral on the left-hand side is expressed by an elliptic function, and then
$\Theta(\sigma)$ 
is expressed using its inverse function as
\begin{equation}
    \Theta(\sigma) = \frac12 \left[
        \pi-4\,\mathrm{am\left(\left.
        (1-2\sigma) F\left(\left.\frac\pi4\right|2\right)
        \right| 2\right)}
    \right]\,
    \label{Thetaexact}
\end{equation}
where $F$ is the elliptic integral of the first kind and the function ``$\mathrm{am}$'' is the amplitude of the Jacobi elliptic functions.
In this expression, we fixed the integration constants $c_0, c_1$ by the boundary conditions  $\Theta(0) = 0$ and $\Theta(\pi) = \pi$.
$\dot X$ and $\dot Z$ are expressed in terms of $\sigma$ by Eqs.~\eqref{Thetadef} and \eqref{Thetaexact}, and $X(\sigma), Z(\sigma)$ are given by their integral.


\item \textit{Neumann boundary at $z=z_0$}: 

For the Neumann boundary condition \eqref{Neumann-BC_WL-z0}, we immediately find an exact solution given by 
\begin{equation}
    X(\sigma)=\text{const.}\,,\qquad
    Z(\sigma) = z_0  \, \sigma\,. \qquad
    (\sigma \in [0,1])
\end{equation}

\end{itemize}

\subsection{Numerical method}

We study how to numerically solve the equation of motion~\eqref{EoM_string-R} under the boundary conditions summarized in section~\ref{sec:WL-BC}.

We use the PINN technique as the numerical solver, with which the boundary value problem can be implemented in a straightforward manner.
We express $R(\theta)$ by a neural network, and conduct machine learning using the following loss function:
\begin{equation}
    \text{Loss} = 
    \frac{1}{N_\text{int}}\sum_{\epsilon \leq \theta \leq \pi/2}\left|\text{ODE loss} \, (\theta)\right|^2
    + |R(\theta=0)- L/2|^2
    + |R'(\theta=\pi/2)|^2\,
\end{equation}
where the ODE loss is the left-hand side of the equation of motion~\eqref{EoM_string-R} itself. The ODE loss is evaluated on random sample points $\theta \in [\epsilon,\pi/2]$, where we introduced a small cutoff $\epsilon$ to avoid a singularity of the equation of motion~\eqref{EoM_string-R} at $\theta=0$.
For the numerical calculations, we have used $N_\text{int} = 100$ sample points to evaluate the ODE loss term.
The second and the third terms enforce the boundary conditions~\eqref{Dirichlet-BC_WL} and \eqref{Neumann-BC_WL}.

\begin{figure}[htbp]
    \centering
    \subfigure[$R(\theta)$]{\includegraphics[width=0.4\linewidth]{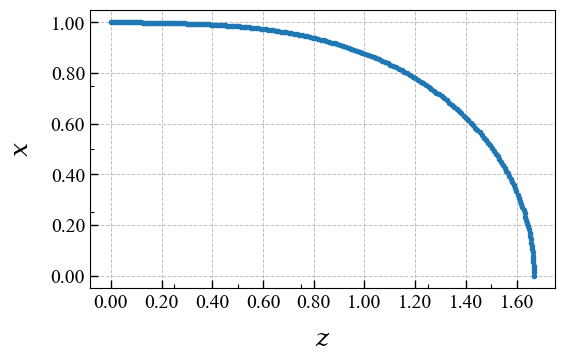}}
    \subfigure[Loss history]{\includegraphics[width=0.4\linewidth]{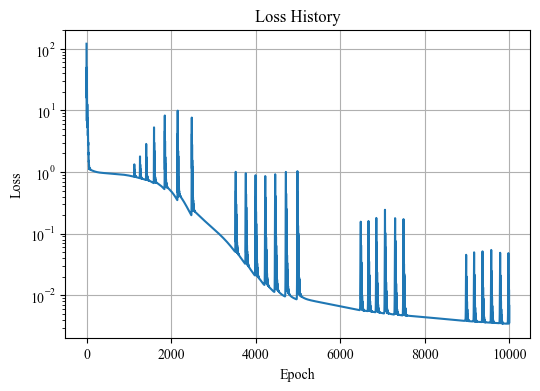}}
    \caption{(a) Wilson loop with $L=2$ obtained by the PINN method. (b) Loss history of the PINN calculation.}
    \label{fig:Wilson-loop_standard}
\end{figure}

In figure~\ref{fig:Wilson-loop_standard}, we show a numerical result of the PINN calculation to find $R(\theta)$ for $L = 2$. We have chosen $\epsilon = 10^{-3}$ for the numerical calculation, 
and used Adam optimizer with the learning rate $\eta = 10^{-3} \times 0.5^{\lfloor(\text{epochs})/2500\rfloor}$.
The location of the Wilson loop at the AdS boundary and also the $z$ axis at the end of the training are given by $R(\theta=0) = 1.00$ and $R(\theta=\pi/2) = 1.67$, which coincides with the analytic solution given in section~\ref{sec:WL-exact} within the numerical accuracy.

\subsection{Introducing Neumann boundary}

We summarize the numerical scheme to find $R(\theta)$ when we introduce a Neumann boundary at $z=z_0$.
We can impose the boundary conditions in the ``soft'' and ``hard'' senses, where the loss terms enforce the boundary conditions in the former, while they are encoded at the formulation level in the latter.\footnote{See e.g.\ \cite{doi:10.1061/(ASCE)EM.1943-7889.0001947,jin2020unsupervised,PhysRevE.105.065305,Luna:2022rql,Luna:2024spo} for the soft (weak) and hard enforcement of the boundary conditions in PINN.}
The numerical code becomes more complicated in enforcing the boundary conditions in the ``hard'' sense. Still, it has the advantage that a numerical solution with better accuracy can be obtained within fewer training epochs.

\begin{enumerate}
    \item \textit{Soft enforcing}:

    \begin{enumerate}
    \item \textit{Truncated loss function}

    One method to impose the Neumann condition~\eqref{Neumann-BC_WL-z0_2} is to generalize the loss function to
    \begin{equation}
    \begin{aligned}
        \text{Loss} &= 
        \frac{1}{N_\text{int}}
        \sum_{\epsilon' \leq R(\theta)\cos\theta \leq z_0}
        \left|\text{ODE loss}\right|^2
        + c_1 \sum_{z_0 \leq R(\theta)\cos\theta \leq z_1}
        \left|R'(\theta) \cos\theta - R(\theta) \sin\theta\right|^2
        \\
        &\quad
    + c_2 \left(|R(\theta=0)- L/2|^2
    + |R(\theta=\pi/2) - z_2|^2
    \right)\,
    \end{aligned}
    \label{loss-WL-Neumann}
    \end{equation}
    where  $\epsilon'$ is a small constant, and $c_1, c_2$ are constant hyper-parameters which we make typically large.
    The ODE loss is the equation of motion~\eqref{EoM_string-R} as before, and the second term in the loss function is the Neumann BC loss, which is nothing but the left-hand side of the boundary condition equation~\eqref{Neumann-BC_WL-z0_2}.
    $z_1, z_2$ are constants satisfying $z_0 < z_1 < z_2$, which are chosen to guarantee that the Neumann condition~\eqref{Neumann-BC_WL-z0_2} is satisfied at $z = z_0$.
    In our numerical code, we chose $c_1=1$ and $c_2=10$.
    We used Adam optimizer with the learning rate $\eta = 10^{-3} \times 0.5^{\lfloor(\text{epochs})/5000\rfloor}$ to obtain the results below.

    We show a numerical result for this formulation in figure~\ref{fig:WL-soft1a}, in which we set $L=2$ and the Neumann condition is imposed at $z = z_0 = 1$.
    The ODE loss and the Neumann BC loss are activated within $\epsilon<z \leq z_0$ and $z_0 \leq z \leq z_1$, respectively, where we set $z_1 = 1.05 \times z_0$.
    The position of the Wilson loop at the $z$ axis $\theta = \pi/2$ is set to $z_2 = 1.5$.
    The part of the Wilson line in $0\leq z \leq z_0 = 1$ corresponds to the solution satisfying the Neumann condition at $z=z_0 = 1$. We can confirm that the Wilson line converges into the correct solution $x = L/2 = \text{constant}$.
    
    \begin{figure}[htbp]
        \centering
        \subfigure[$R(\theta)$]{\includegraphics[width=0.4\linewidth]{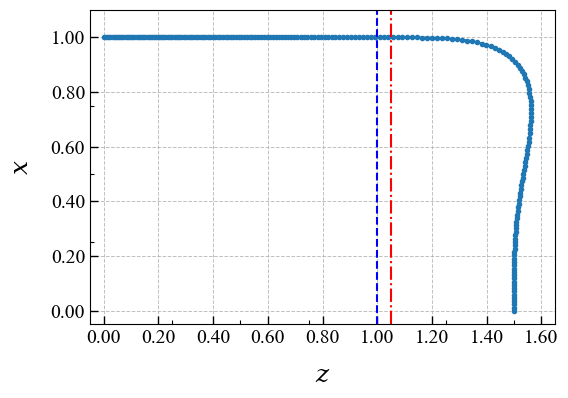}}
        \subfigure[Loss history]{\includegraphics[width=0.4\linewidth]{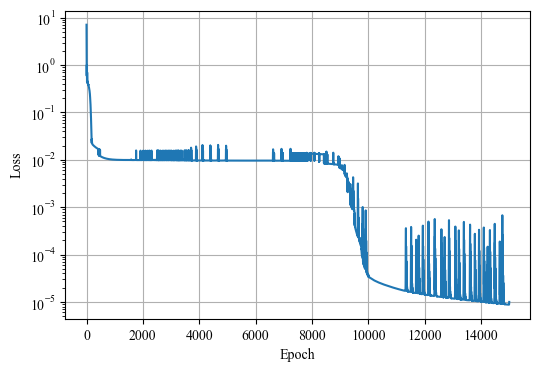}}
        \caption{(a) Wilson loop with $L=2$ obtained by the PINN method with a truncated loss function~\eqref{loss-WL-Neumann}. 
        The blue dashed and red dot-dashed lines correspond to $z=z_0$ and $z_1$, respectively. The Neumann BC loss is activated in the region between these two lines. The curve in $0\leq z \leq z_0 = 1$ describes the Wilson loop with the Neumann conditions imposed at $z=z_0$. (b) Loss history in this case.}
        \label{fig:WL-soft1a}
    \end{figure}

    \item \textit{Moving-boundary scheme}

    An alternative method to impose the Neumann condition~\eqref{Neumann-BC_WL-z0_2} is to promote $\theta_0$ satisfying $R(\theta_0) \sin\theta_0 = z_0$ to a parameter that is optimized in the PINN calculation along with the neural network expressing $R(\theta)$.
    Then, we conduct the training with the loss function defined as
    \begin{equation}
        \text{Loss} 
        =
        \frac{1}{N_\text{int}}
        \sum_{\epsilon \leq \theta \leq \theta_0}
        \left|\text{ODE loss}\right|^2
        + c_1 \left|R'(\theta_0) \cos\theta_0 - R(\theta_0) \sin\theta_0\right|^2
        + c_2 \left|R(\theta=0)- \frac{L}{2} \right|^2 \,.
    \label{loss-WL-Neumann_moving-theta0}
    \end{equation}
    The boundary of the calculation domain $\theta \in [0,\theta_0]$ moves in the training process of the PINN calculation in this formulation.
    If the training proceeds correctly, $\theta_0$ converges to the correct value corresponding to the true solution.

    In figure~\ref{fig:WL-soft1b}, we show a numerical result based on the above numerical scheme for $L=2$ and $z_0 = 1$. We can observe that the numerical solution converges to a correct one $x = L/2 = \text{constant}$ within the numerical accuracy.
       
    \begin{figure}[htbp]
        \centering
        \subfigure[$R(\theta)$]{\includegraphics[width=0.3\linewidth]{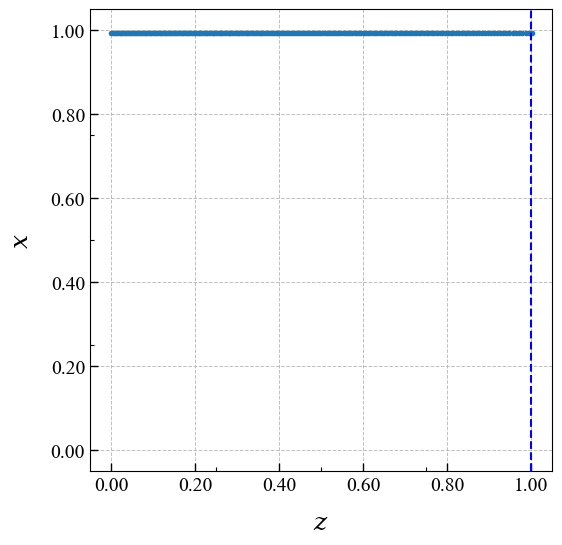}}
        \subfigure[Loss history]{\includegraphics[width=0.4\linewidth]{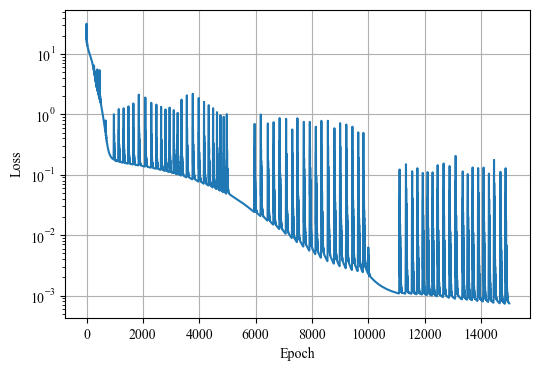}}
        \caption{(a) Wilson loop with $L=2$ obtained by the moving-boundary scheme with the loss function~\eqref{loss-WL-Neumann_moving-theta0}. 
        The Neumann condition is imposed at $z=z_0=1$ (blue dashed line). (b) Loss history in this case.}
        \label{fig:WL-soft1b}
    \end{figure}

    \end{enumerate}
    
    \item \textit{Hard enforcing}

    The boundary condition may be encoded at the level of formulation instead of enforcing it by adding the boundary condition loss terms.
    One way to implement the boundary conditions \eqref{Dirichlet-BC_WL2}, \eqref{Neumann-BC_WL-z0} and \eqref{Neumann-BC_WL-z0_2} is to express $R(\theta)$ as
    \begin{equation}
        R(\theta) = \mathcal{B}(\theta; \theta_0) + \mathcal{E}(\theta; \theta_0)\times f_\text{NN}(\theta)\,,
        \label{R-ansatz_WL}
    \end{equation}
    where $\mathcal{B}$ and $\mathcal{E}$ are the ``baseline'' and ``envelope'' functions defined as
    \begin{align}
        \mathcal{B}(\theta; \theta_0)
        &= \frac{L}{2}\left(1 + \theta^2\right)
         + c_3\, \theta^2 + c_4 \, \theta^4\,,
         \\
         \mathcal{E}(\theta; \theta_0) &= \theta^3 (\theta_0-\theta)^2\,.
    \end{align}
    $\theta=\theta_0$ is the angle at which the Neumann condition~\eqref{Neumann-BC_WL-z0_2} is imposed, and its value is determined by the training in the PINN calculation along with the function $f_\text{NN}(\theta)$.
    The coefficients $c_3, c_4$ are determined so that $\mathcal{B}(\theta; \theta_0)$ satisfies\footnote{The explicit expression of $c_3, c_4$ satisfying \eqref{B-BC_WL} are given by
    \begin{equation*}
    c_3 = -\frac{\frac{1}{2} L \left(\theta_0{}^2+4\right)-4 R_0+R'_0 \theta_0{}}{\theta_0{}^3}\,,
    \qquad
    c_4 = \frac{\frac{1}{2} L \left(\theta_0{}^2+6\right)-6 R_0+2 R'_0 \theta_0{}}{2 \, \theta_0{}^4}\,,
    \end{equation*}
    where $R_0 = z_0 / \sin\theta_0$ and $R'_0 =  z_0 / \cos\theta_0$.}
    \begin{equation}
        \mathcal{B}(\theta = \theta_0; \theta_0) = \frac{z_0}{\sin\theta_0}\,,
        \qquad
        \partial_\theta \mathcal{B}(\theta=\theta_0; \theta_0) = \frac{z_0}{\cos\theta_0}\,.
        \label{B-BC_WL}
    \end{equation}

    The definitions above guarantee that $R(\theta)$ satisfies the boundary conditions \eqref{Dirichlet-BC_WL2}, \eqref{Neumann-BC_WL-z0} and \eqref{Neumann-BC_WL-z0_2} for any $\theta_0$ and $f_\text{NN}(\theta)$ provided that $f_\text{NN}(\theta)$ is finite everywhere.
    Then, we can find a numerical solution by taking the loss function as
    \begin{equation}
    \text{Loss} 
    = 
    \frac{1}{N_\text{int}}\sum_{\epsilon \leq \theta \leq \theta_0}
    \left|\text{ODE loss}\right|^2\,.
    \label{loss-WL-Neumann_hard}
    \end{equation}
    We do not need to include the loss terms corresponding to the boundary conditions since they are automatically satisfied by using the ansatz~\eqref{R-ansatz_WL}.
    For the numerical results shown below, we used $N_\text{int}$ sample points taken randomly in $\epsilon \leq \theta \leq \theta_0$.

    We show the numerical result for this scheme for $L=2$ and $z_0 = 1$. An advantage of this scheme is that the numerical solution converges to a correct one ($x = L/2 = \text{constant}$) within fewer training epochs; Figure~\ref{fig:WL-hard_loss} shows that the loss function becomes smaller within fewer epochs compared to the other cases shown in figures~\ref{fig:WL-soft1a} and \ref{fig:WL-soft1b}.
    The drawback is that the ansatz~\eqref{R-ansatz_WL} and the numerical code based on it become more complicated than those in the other schemes.

    \begin{figure}[htbp]
        \centering
        \subfigure[$R(\theta)$]{\includegraphics[width=0.28\linewidth]{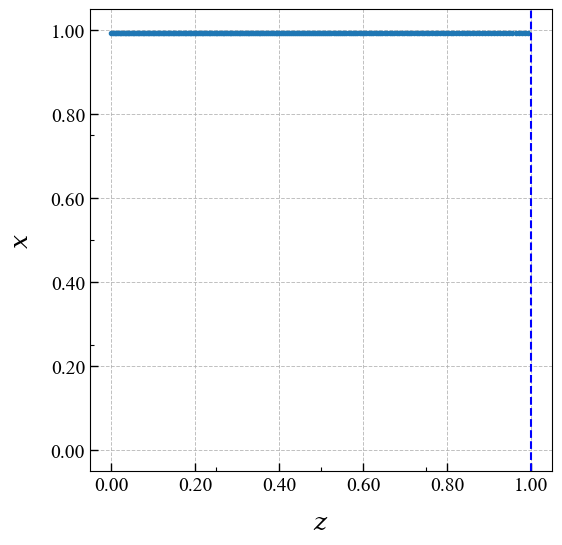}}
        \subfigure[Loss history]{\includegraphics[width=0.4\linewidth]{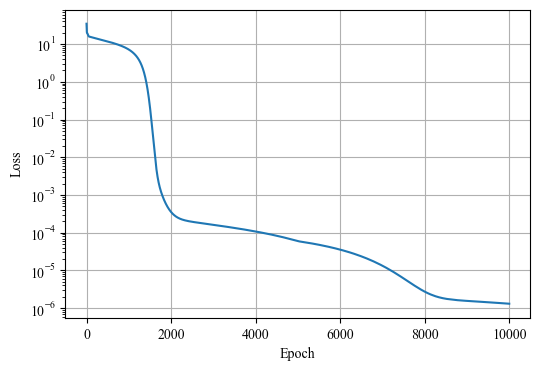} \label{fig:WL-hard_loss}}
        \caption{(a) Wilson loop with $L=2$ obtained by the hard-enforcing scheme with the loss function~\eqref{loss-WL-Neumann_hard}. 
        The Neumann condition is imposed at $z=z_0=1$ (blue dashed line). (b) Loss history in this case.}
        \label{fig:WL-hard}
    \end{figure}
    
 \end{enumerate}

\section{Two-dimensional minimal surface in curved space}
\label{sec:minimal-surface}

We turn to the next simplest case of the two-dimensional minimal surface in the AdS spacetime. As an illustration of the capability of the PINN method, we work on the following two cases:
1.\ standard minimal surface in AdS, and 2.\ minimal surface with an additional Neumann boundary. In the second case, we impose the Neumann boundary condition at a fixed radial position $r=r_0$.\footnote{The imposition of the additional boundary amounts, in string theory,  to the introduction of a D-brane parallel to the AdS boundary. In the AdS/CFT correspondence, this corresponds to considering a Coulomb phase of the ${\cal N}=4$ supersymmetric Yang-Mills theory, or introducing a quark hypermultiplet to the theory. }
When this problem is formulated using polar coordinates, it becomes a boundary-value problem with a moving boundary. Such a problem can be naturally formulated and efficiently solved using the PINN technique, as shown below.

\subsection{Setup}
\label{sec:minimal-surface_setup}

We consider a two-dimensional minimal surface in an Euclidean AdS$_3$ spacetime with a metric
\begin{equation}
 ds^2 = \frac{1}{r^2}\left(dy_1^2 + dy_2^2 + dr^2\right)~.
\end{equation}
We assume that the edge of the minimal surface attaches to the AdS boundary along a prescribed curve on it. 

To describe the minimal surface shape, we use the spherical coordinates defined by
\begin{equation}
    y_1 = R(\theta,\phi) \sin\theta\cos\phi\,,\quad
    y_2 = R(\theta,\phi) \sin\theta\sin\phi\,,\quad
    r = R(\theta,\phi) \cos\theta\,.
    \label{sph-coords_R}
\end{equation}
Then, the Lagrangian describing this surface is given by
\begin{equation}
    L = \frac{1}{R \cos^2\theta}
    \sqrt{
    \left( R^2  + \left(\partial_\theta R\right)^2\right)\sin^2\theta
    + \left(\partial_\phi R\right)^2
    }\,.
    \label{Lag_R}
\end{equation}
and the Euler-Lagrange equation for this Lagrangian is given by an elliptic PDE for $R(\theta,\phi)$ whose explicit expression is not illuminating.

We make the following two assumptions: the minimal surface is symmetric with respect to the $y_1$ and $y_2$ axes; the surface is described by~\eqref{sph-coords_R} with a single-valued function $R(\theta,\phi)$.
The first of these is tantamount to assuming
\begin{equation}
    \partial_\phi R = 0
    \qquad
    \left( \phi=0, \, \frac{\pi}{2}; ~\theta \in [0,\pi/2] \right)\,.
    \label{Neumann-phi}
\end{equation}
Also, thanks to the reflection symmetry with respect to the $y_1, y_2$ axes we assumed, we may take the calculation domain as
\begin{equation}
    (\theta,\phi) \in [0, \pi/2]\times[0,\pi/2]\,,
    \label{R_domain}
\end{equation}
where $\theta = 0$ and $\pi/2$ correspond to the rotational axis of the spherical coordinates and the AdS boundary, respectively.

We solve the Euler-Lagrange equation obtained from \eqref{Lag_R} under the following boundary conditions.
\begin{itemize}
    \item \textit{Boundary conditions on the AdS boundary}:
    
We require that the edge of the surface at the AdS boundary coincides with a curve prescribed by a given function $R_\text{bdy}(\phi)$, that is,
\begin{equation}
    R(\theta = \pi/2,\phi) = R_\text{bdy}(\phi)
    \qquad\left(\phi\in[0,\pi/2]\right)
    \,.
    \label{R-BC_Dirichlet}
\end{equation}
In other words, we construct a minimal surface with a Dirichlet condition \eqref{R-BC_Dirichlet} imposed.
We assume that $R_\text{bdy}(\phi)$ is given by a smooth function for simplicity.
Also, the symmetry assumption~\eqref{Neumann-phi} requires
\begin{equation}
    \partial_\phi R_\text{bdy}(\phi) = 0
    \qquad
    \left(\phi = 0, \, \frac{\pi}{2}\right)\,.
\end{equation}

By constructing a series solution of the equation of motion near the AdS boundary, we can show that a regular solution obeying the above boundary condition must satisfy
\begin{equation}
    \partial_\theta R(\theta=0,\phi)=0\qquad\left(\phi\in[0,\pi/2]\right)\,.
    \label{R-BC_Neumann-AdSbdy}
\end{equation}

\item \textit{Neumann conditions at $\phi = 0, \pi/2$}:

We assume that 
$R(\theta,\phi)$ and $R_\text{bdy}(\phi)$ satisfies the Neumann condition in the $\phi$ direction
\begin{equation}
\partial_\phi R(\theta,\phi) \,\Bigr|_{\phi=0,\pi/2}
    =0
    \quad (\theta \in [0,\pi/2])\,,
    \qquad
    \partial_\phi R_\text{bdy}(\phi)\,\Bigr|_{\phi=0,\pi/2} = 0
    \label{R-BC_Neumann}
\end{equation}
to be compatible with the reflection symmetry at the $y_1,y_2$ axes.

\item \textit{Regularity condition at the axis $\theta=0$}:

It is straightforward to see that the regularity condition of the surface at the axis $\theta = 0$ under the above assumptions boils down to the Neumann condition in the $\theta$ direction, that is,
\begin{equation}
    \partial_\theta R(\theta=0,\phi) = 0
    \qquad
    \left(\phi\in[0,\pi/2]\right)
    \,.
    \label{R-BC_regularity}
\end{equation}

\end{itemize}

Equations~\eqref{Neumann-phi}, \eqref{R-BC_Dirichlet}, \eqref{R-BC_Neumann}, and \eqref{R-BC_regularity} comprise the boundary conditions to solve the equation of motion obtained from the Lagrangian~\eqref{Lag_R} as a boundary-value problem.
The condition~\eqref{R-BC_Neumann-AdSbdy} should be satisfied automatically if we construct a regular solution satisfying the equation of motion. For the convenience of the numerical calculation, however, we explicitly enforce this condition by the method described below.

\subsection{Numerical method}
\label{sec:manybody-method}

We employ the PINN technique to construct numerical solutions in the setting introduced above.
Since the equation of motion possesses singularities at 
the axis $\theta=0$ and the AdS boundary $\theta=\pi/2$, we need to treat them carefully to stabilize the numerical computation.
For this purpose, we express $R(\theta,\phi)$ without loss of generality as\footnote{See e.g.\ \cite{TSENG2023112359,hu2024solving,Cayuso:2024jau} for earlier studies on PINN with various singularities.}
\begin{equation}
    R(\theta,\phi) = 
    \mathcal{B}(\theta, R_\text{bdy}(\phi),R_\text{tip})
    + \mathcal{E}(\theta) \times f_\text{NN}(\theta,\phi)\,,
    \label{R_PINN-ansatz}
\end{equation}
where the functions $\mathcal{B}, \mathcal{E}$ are defined as
\begin{align}
    \mathcal{B}(\theta, R_\text{bdy}(\phi),R_\text{tip}) &:=
    R_\text{tip} 
    + \left(R_\text{bdy}(\phi) - R_\text{tip}\right) \left(\frac{\theta}{\pi}\right)^2 \left( 12 - 16 \frac\theta\pi\right)\,,
    \\
    \mathcal{E}(\theta) &:= \theta^2 \left(\theta-\frac\pi2\right)^2 \,,
    \label{eq:Rtip}
\end{align}
and $f_\text{NN}(\theta,\phi)$ is a function expressed by a neural network.
%
The network consists of
four hidden layers, each with $50$ neurons.  All hidden layers use the $\tanh$ activation function.
Here, $R_\text{tip}$ is the height of the surface at the axis $\theta = 0$, that is, $R(\theta=0,\phi) = R_\text{tip}$.
Its value can be fixed only after solving the equation of motion under the boundary conditions described above.
As we explain below, we optimize $f_NN(\theta,\phi)$ and $R_\text{tip}$ simultaneously by the PINN calculation.
The function form of the ``baseline'' function $\mathcal{B}$ and the ``envelope'' function $\mathcal{E}$ were determined so that, as long as $f_\text{NN}(\theta,\phi)$ is finite everywhere, $R(\theta,\phi)$ given by~\eqref{R_PINN-ansatz} behaves as
\begin{align}
    R(\theta,\phi) &= R_\text{tip} + \theta^2 \times f(\phi) 
    + \cdots
    &\left(\theta \simeq 0\right) ~
    \\
    R(\theta,\phi) &= R_\text{bdy}(\phi) + \left(\theta - \frac\pi2\right)^2 \times g(\phi) + \cdots
    &\left(\theta \simeq \frac\pi2\right)
    \label{BC-phi_AdS}
\end{align}
for some functions $f(\phi), g(\phi)$, that is, the boundary conditions summarized in section~\ref{sec:minimal-surface_setup} are enforced automatically as long as $f_\text{NN}(\theta,\phi)$ is finite.\footnote{The expression \eqref{BC-phi_AdS} can be also obtained by solving the Euler-Lagrange equation order by order in small $(\theta - \tfrac{\pi}{2})$ at $\theta = \tfrac{\pi}{2}$. The solution ansatz \eqref{R_PINN-ansatz} guarantees that the behavior of $R(\theta,\phi)$ coincides with that of the correct (series) solution at $\theta \simeq \tfrac{\pi}{2}$.
By this construction, not only the boundary condition~\eqref{R-BC_Dirichlet}, but also the Euler-Lagrange equation is guaranteed to be satisfied near $\theta = \tfrac{\pi}{2}$.}
An advantage of this method is that we do not need to introduce a term corresponding to these boundary conditions into the loss function, and then the numerical solution converges to the correct one in fewer training epochs.

In the PINN calculation, we optimize the function $f_\text{NN}(\theta,\phi)$ as well as the constant $R_\text{tip}$ to minimize the following loss function given by
\begin{equation}
    \text{Loss} = \frac{1}{N_\text{int}}\sum_{\theta,\phi}\left|\text{PDE loss} \, (\theta,\phi)\right|^2
    + \frac{1}{N_\text{bc}}\sum_{\theta} \left(
     \left|\partial_\phi f_\text{NN}(\theta,\phi=0)\right|^2
    +\left|\partial_\phi f_\text{NN}(\theta,\phi=\pi/2)\right|^2
    \right)\,,
\end{equation}
where the summations are evaluated at random points on the numerical domain and its boundary.
The second term in this loss function, along with assumptions on $R_\text{bdy}(\phi)$, guarantees that $R(\theta,\phi)$ satisfies the Neumann condition $\partial_\phi R(\theta, \phi=0,\pi/2) = 0$.

We train with Adam optimizer with a cosine–annealing schedule. 
The learning rate at the epoch $t$ is given by
\[
\eta(t)=\eta_{\min}+\frac{1}{2}\bigl(\eta_{\max}-\eta_{\min}\bigr)
\bigl[1+\cos\bigl(\pi\,t/T_{\max}\bigr)\bigr],\qquad t=0,\dots,T_{\max}-1\,,
\]
which starts from $\eta = \eta_{\max}$ to $\eta_{\min}$ smoothly toward the end of the training at $T_{\max}$ epochs.
This provides large exploratory steps early on and a smooth, non-oscillatory decay to a fine-tuning regime, which is beneficial for PDE-constrained training involving second derivatives.
In our runs, we set $\eta_{\max}=10^{-3}$, $\eta_{\min}=10^{-6}$, and we vary $T_{\max}$ for each problem.

\subsection{Standard minimal surface}
\label{sec:standard-minimal-surface}

\begin{figure}[htbp]
    \centering
    \subfigure[$r(y_1,y_2)$]{\includegraphics[width=7cm]{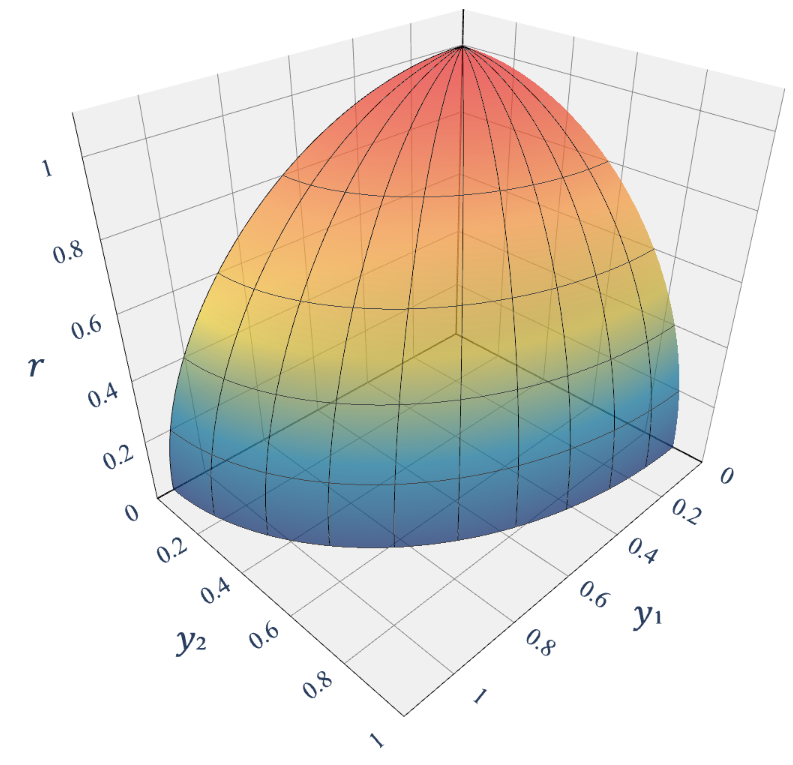}}
    \qquad
    \subfigure[Loss history]{\includegraphics[width=7cm]{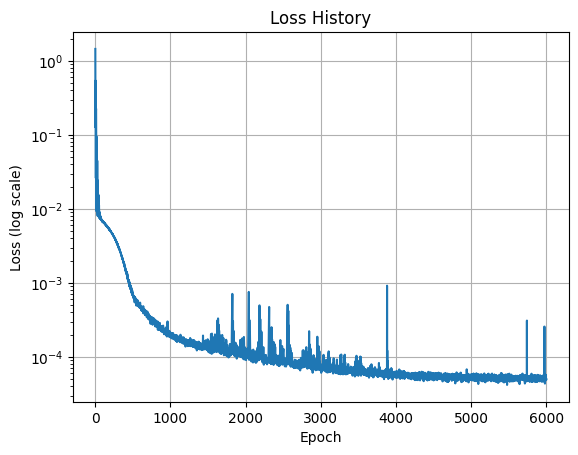}}
    \caption{Panel (a): profile of $R(\theta,\phi)$ for $R_\text{bdy}(\phi) = 1+0.1\times \cos(2\phi)$ shown with respect to $(y_1, y_2)$.
    The grid lines correspond to constant $r$ and $\phi$.
    Panel (b): loss history.}
    \label{fig:Dome}
\end{figure}

In figure~\ref{fig:Dome}, we show the numerical solution of the surface $R(\theta,\phi)$ whose edge is located at
\begin{equation}
    R(\theta=\pi/2,\phi) = R_\text{bdy}(\phi) = 1 + 0.1 \times \cos(2\phi)\,.
\end{equation}
For this $R_\text{bdy}(\phi)$, the tip of the surface is at $R_\text{tip} = 1.002$ at the end of the training.
In this calculation, we have taken $N_\text{int}=5000$ and $N_\text{bc}=500$, and the sample points are taken uniformly and randomly.
At the end of the training (6000 epochs), the total loss is $5.0\times 10^{-5}$, whose $\sim 60\%$ is comprised of the PDE loss and $\sim 40\%$ is of the boundary condition loss.

When $R_\text{bdy}(\phi)$ is a constant independent of $\phi$, we can confirm that the numerical solution $R(\theta,\phi)$ converges to the exact solution $R(\theta,\phi) = R_\text{bdy}$ within the numerical accuracy.


\subsection{Introducing Neumann boundary}
\label{sec:minimal-surface_N}



As an extension of the standard case, we consider a minimal surface in the AdS spacetime with an additional boundary at constant AdS radial value $r=r_0$. Such a problem naturally arises when, e.g., the AdS spacetime is cut off or when an additional D-brane is inserted.

At $r=r_0$, we assume that the minimal surface satisfies the Neumann boundary condition given by
\begin{align}
    \partial_\theta \left(R(\theta,\phi)\sin\theta\right)\bigr|_{R \cos\theta = r_0} = 0
    \qquad&\Leftrightarrow\qquad
    \partial_\theta R + R\cot\theta
    \,\bigr|_{R \cos\theta = r_0}
    = 0
    \\
    &\Leftrightarrow\qquad
    \partial_\theta R + \frac{r_0}{\sin\theta}
    \,\Bigr|_{R \cos\theta = r_0}
    = 0
    \label{Neumann_sph}
\end{align}
This condition follows from the requirement that the surface has a minimal area.

The numerical method for the standard minimal surface can be generalized to incorporate the boundary condition~\eqref{Neumann_sph}.
We propose the following two ways to realize it.
    
    \subsubsection{Soft enforcing}
    \label{sec:minimal-surface_soft}

\begin{figure}
        \centering
        \includegraphics[width=0.5\linewidth]{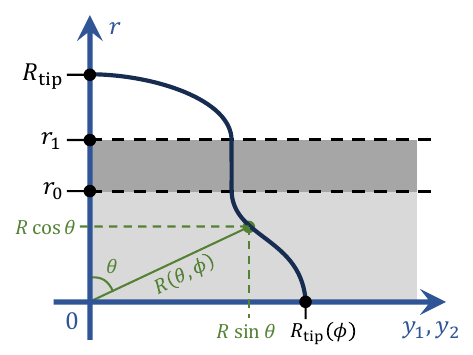}
        \caption{ A schematic for the structure of the loss function to introduce
        a Neumann boundary at $r=r_0$.
        The PDE loss is given by the Euler-Lagrange equation in $0\leq r \leq r_0$ (light gray region), while it is switched to the equation for the Neumann boundary condition in $r_0 \leq r \leq r_1$ (dark gray region).}
        \label{fig:manybody-Neumann-scheme}
    \end{figure}

    In the first method, we generalize the framework used in section~\ref{sec:manybody-method} as follows.
    We express $R(\theta,\phi)$ as
    \begin{equation}
    R(\theta,\phi) = 
    \mathcal{B}(\theta, R_\text{bdy}(\phi),R_\text{tip})
    + \mathcal{E}(\theta) \times f_\text{NN}(\theta,\phi)\,,
    \label{R_PINN-ansatz_Neumann}
    \end{equation}
    where we optimize $f_\text{NN}(\theta,\phi)$ by the PINN calculation as before, while we fix $R_\text{tip}$ to be a constant as described below.
    We also generalize the loss function as
    \begin{equation}
    \begin{aligned}
        \text{Loss} &=
        \frac{1}{N_\text{int}}
        \sum_{\theta,\phi} 
        \left[
        \tophat{0}{r_0}(R\cos\theta)\left|\text{PDE loss}\right|^2
        + \tophat{r_0}{r_1}(R\cos\theta)\left|\text{Neumann BC loss}\right|^2
        \right]
        \\
        &\quad
        + \frac{1}{N_\text{bc}} 
         \sum_{\theta} \left(
         \left|\partial_\phi R(\theta,\phi=0)\right|^2
        +\left|\partial_\phi R(\theta,\phi=\pi/2)\right|^2
        \right)\,,
    \end{aligned}
    \label{loss_minimal-surface-Neumann}
    \end{equation}
    where $\tophat{a}{b}(x)$ is a top-hat window function defined as
    \begin{equation}
        \tophat{a}{b}(x) :=
        \begin{cases}
            1 & (a \leq x \leq b)\\
            0 & (\text{else})
        \end{cases} \,.
    \end{equation}
    The ``PDE loss'' is the same as before, and the ``Neumann BC loss'' is given by the left-hand side of \eqref{Neumann_sph}.
    We set $r_0$, $r_1$ and $R_\text{tip}$ so that they satisfy $r_0 < r_1 < R_\text{tip}$ and are separated by some width.
    See figure~\ref{fig:manybody-Neumann-scheme} for an illustration of the structure of the loss function.
    
    If the loss function~\eqref{loss_minimal-surface-Neumann} becomes zero as a result of the PINN calculation, the part of the surface for $r\in[0, r_0]$ coincides with a minimal surface satisfying the Neumann condition at $r=r_0$.
    Advantages of this method are that the numerical domain to solve the equations is fixed to $(\theta,\phi)\in [0,\pi/2]\times[0,\pi/2]$ and also that we may use the numerical code for the previous case almost unaltered.

    In~\eqref{loss_minimal-surface-Neumann}, we may take the width of the region to impose the Neumann condition~\eqref{Neumann_sph} infinitely thin by making the coefficient for this term in the loss function large in principle. Instead, we take the width finite while keeping the coefficient to unity, and also fix $R_\text{tip}$ larger than $r_1$ to make sure that the solution of the PINN robustly converges into a correct solution satisfying the Neumann condition at $r=r_0$.

\begin{figure}[htbp]
    \centering
    \subfigure[$r(y_1,y_2)$]{\includegraphics[width=7cm]{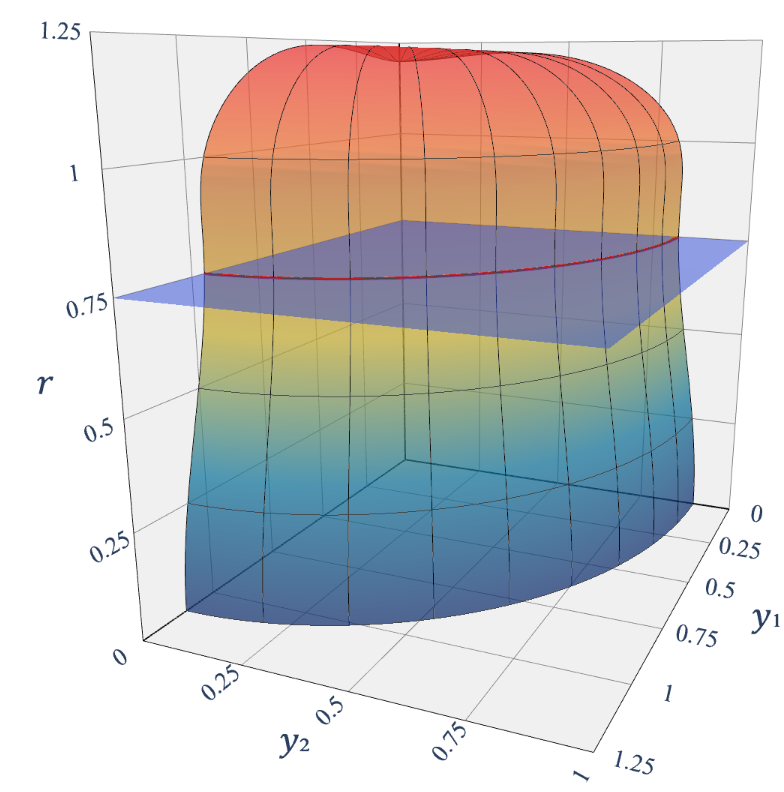}  \label{fig:Dome-Neumann}}
    \qquad
    \subfigure[Loss history]{\includegraphics[width=7cm]{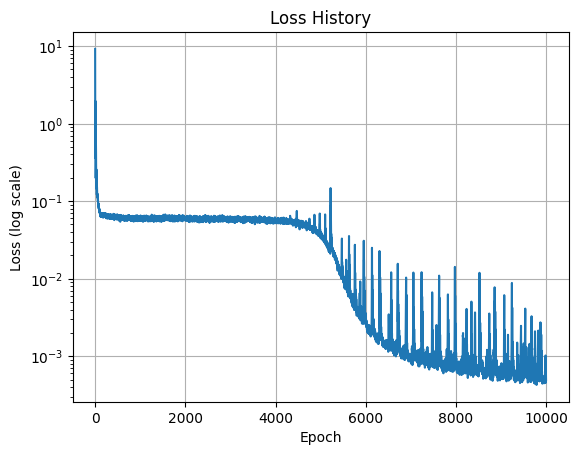}\label{fig:loss_round-soft}}
    \caption{Panel (a): profile of $R(\theta,\phi)$ for $R_\text{bdy}(\phi) = 1+0.1\times \cos(2\phi)$ when 
    the Neumann boundary is located
    at $r=0.75$ (blue plane). The part of the surface for $r\in[0,0.75]$ represents the minimal surface satisfying the Neumann condition at $r = 0.75$.
    Panel (b): loss history.    }
    \label{fig:Dome-Neumann-and-loss}
    \end{figure}

    In figure~\ref{fig:Dome-Neumann}, we show a numerical solution of a minimal surface when the Neumann boundary is located at $r = r_0 = 0.75$ and the boundary shape is given by $R_\text{bdy}(\phi) = 1+0.1\times \cos(2\phi)$.
    We conducted this calculation with the sampling points the same as before: $N_\text{int} = 5000$, $N_\text{bc}=500$, and the sampling points are distributed uniformly and randomly.
    For this calculation, we took $r_1 = 1.2\times r_0$ and $R_\text{tip} = 1.4 \times r_0$.
    The part of this solution for $r\in[0,0.75]$ describes the minimal surface satisfying the Neumann condition at $r=0.75$.

    Figure~\ref{fig:loss_round-soft} shows the loss history for this calculation.
    The loss history develops a peculiar plateau-like structure near the beginning of the training. In this regime, the surface takes a shape similar to that of the standard minimal surface without the Neumann boundary presented in section~\ref{sec:standard-minimal-surface}. After this regime, the total loss begins to decrease again, and the surface shape given in figure~\ref{fig:Dome-Neumann} is obtained at an asymptotically late time in this regime.
    
At the end of the training ($10^4$ epoch), the total loss is \(7.91\times 10^{-4}\), decomposed as
\((\text{PDE Loss})=5.54\times 10^{-4}\) (70.09\%),
\((\text{Neumann BC loss})=1.74\times 10^{-4}\) (22.02\%), and
\((\text{$\partial_\phi R$ loss})=6.23\times 10^{-5}\) (7.89\%).
This indicates that convergence is chiefly limited by the interior PDE residual, 
and the Neumann boundary condition contributes subdominantly yet non-negligibly within \(r\in[r_0,r_1]\).
The \(\phi\)-edge Neumann boundary condition is well enforced.

    \subsubsection{Hard enforcing}
    \label{sec:minimal-surface_hard}

    The second method to impose the Neumann condition at $r=r_0$ is to encode this condition at the level of the formulation. By doing this, the boundary condition is automatically satisfied, and then the training ends within fewer epochs than the previous method. A drawback is that the formulation may become more complicated and less amenable to more general boundary conditions.

    In this method, we take the calculation domain as
    \begin{equation}
        (\theta,\phi) \in [\theta_0(\phi),\pi/2] \times [0,2\pi]\,
    \end{equation}
    where we impose the Neumann condition at $\theta = \theta_0(\phi)$, that is,
    \begin{equation}
        R\bigl(\theta_0(\phi),\phi\bigr) \cos\theta_0(\phi) =  r_0\,,
        \qquad
        \partial_\theta R\bigl(\theta_0(\phi),\phi\bigr)
        + \frac{r_0}{\sin\theta_0(\phi)}  
        = 0\,.
        \label{R_PINN-BC_Neumann-hard}
    \end{equation}
    The position of the boundary $\theta = \theta_0(\phi)$ is fixed only after obtaining a solution in the entire calculation domain. Below, we express $\theta_0(\phi)$ by a neural network along with $R(\theta,\phi)$ and train them simultaneously by the PINN method.
    To guarantee the smoothness of the section at $\theta=\theta_0(\phi)$, we need to impose the Neumann boundary condition to $\theta_0(\phi)$ at $\phi=0,\pi/2$.

    To enforce the boundary condition
    we begin with expressing $R(\theta,\phi)$ as follows:
    \begin{equation}
    R(\theta,\phi) = 
    \mathcal{B}\bigl(\theta, R_\text{bdy}(\phi),r_0,\theta_0(\phi)\bigr)
    + \mathcal{E}\bigl(\theta,\theta_0(\phi)\bigr) \times f_\text{NN}(\theta,\phi)\,,
    \label{R_PINN-ansatz_Neumann-hard}
    \end{equation}
    As the baseline function $\mathcal{B}$ and the envelope function $\mathcal{E}$, we may take an arbitrary function satisfying the following conditions:
    \begin{align}
        &\hphantom{\partial_\theta}\mathcal{B}\Bigl(\theta=\frac{\pi}{2}, R_\text{bdy}(\phi),r_0,\theta_0(\phi) \Bigr) = R_\text{bdy}(\phi)\,, \label{Bcond1} \\
        &\partial_\theta \mathcal{B}\Bigl(\theta=\frac{\pi}{2}, R_\text{bdy}(\phi),r_0,\theta_0(\phi) \Bigr) = 0\,, \label{Bcond2}\\
        &\hphantom{\partial_\theta}\mathcal{B}\bigl(\theta=\theta_0(\phi), R_\text{bdy}(\phi),r_0,\theta_0(\phi) \bigr) \cos\theta_0(\phi) = r_0\,,
        \label{Bcond3}\\
        &\partial_\theta \mathcal{B}\bigl(\theta=\theta_0(\phi), R_\text{bdy}(\phi),r_0,\theta_0(\phi) \bigr)
        + \frac{r_0}{\sin\theta_0(\phi)}
        = 0\,,
        \label{Bcond4}
    \end{align}
    \begin{equation}
        \mathcal{E}\Bigl(\theta \sim \frac{\pi}{2},\theta_0(\phi)\Bigr) \sim \left(\theta - \frac{\pi}{2}\right)^2\,,
        \qquad
        \mathcal{E}\bigl(\theta \sim \theta_0(\phi),\theta_0(\phi)\bigr) \sim \bigl(\theta - \theta_0(\phi)\bigr)^2\,.    
    \end{equation}
    For $\mathcal{B}, \mathcal{E}$ satisfying the above conditions, $R(\theta,\phi)$ defined by \eqref{R_PINN-ansatz_Neumann-hard} satisfies the boundary condition \eqref{R-BC_Dirichlet}, \eqref{R-BC_Neumann-AdSbdy} at the AdS boundary and also the condition \eqref{R_PINN-BC_Neumann-hard} at the additional boundary at $r=r_0$ provided that $f_\text{NN}(\theta,\phi)$ in \eqref{R_PINN-ansatz_Neumann-hard} is finite everywhere.
    This prescription makes the numerical code more robust against possible numerical errors generated near the AdS boundary, where the equation of motion becomes singular. Also, the PINN calculation converges more quickly because the profile of $f_\text{NN}(\theta, \phi)$ is closer to a constant profile with less gradient compared to $R(\theta,\phi)$.

    The baseline function $\mathcal{E}$ and the envelope function $\mathcal{E}$ are arbitrary as long as they satisfy the conditions shown above. In our study, we use the following functions given by polynomials:
    \begin{align}
        \mathcal{B}\bigl(\theta, R_\text{bdy}(\phi),r_0,\theta_0(\phi)\bigr)
        &= C_0 + C_1 \, \theta + C_2 \, \theta^2 + C_3 \, \theta^3\,,
        \label{Bpol}
        \\
        \mathcal{E}\bigl(\theta,\theta_0(\phi)\bigr) &=
        \bigl(\theta - \theta_0(\phi)\bigr)^2
        \left(\theta - \frac{\pi}{2} \right)^2\,,
    \end{align}
    where $C_0, C_1, C_2, C_3$ are functions of $R_\text{bdy}(\phi)$ and $\theta_0(\phi)$ that are determined by imposing the conditions~\eqref{Bcond1}--\eqref{Bcond4}.\footnote{Explicit form of \eqref{Bpol} after imposing the conditions~\eqref{Bcond1}--\eqref{Bcond4} is given by
    \begin{multline*}
    \mathcal{B}\bigl(\theta, R_\text{bdy}(\phi),r_0,\theta_0(\phi)\bigr)
    =
        \frac{1}{(\pi -2 \theta_0)^3}
        \Bigl\{
   -4 R_\text{bdy} (\theta_0-\theta )^2 (4 \theta +2 \theta_0-3 \pi )
   \\
   +(\pi -2 \theta )^2 r_0 \bigl[
   (\pi -2 \theta_0) (\theta_0-\theta ) \csc (\theta_0)+(4 \theta -6 \theta_0+\pi )
   \sec (\theta_0)
   \bigr]
   \Bigr\}\,.
    \end{multline*}
    }
 
    Since the boundary conditions at $\theta = 0$ and $\theta=\theta_0(\phi)$ are automatically satisfied in this prescription, we do not need to introduce terms corresponding to them into the loss function. Then, the loss function in this case is given by
    \begin{align}
        \text{Loss} &=
        \frac{1}{N_\text{int}}\sum_{\theta,\phi} \left|\text{PDE loss}\right|^2
        + \frac{1}{N_\text{bc}}\sum_{\theta} 10\left(
         \left|\partial_\phi R(\theta,\phi=0)\right|^2
        +\left|\partial_\phi R(\theta,\phi=\pi/2)\right|^2
        \right)
        \notag \\
        &\quad
        + 10 \left[
            |\partial_\phi \theta_0(\phi=0)|^2
            + |\partial_\phi \theta_0(\phi=\pi/2)|^2
        \right]
        \,,
    \label{loss_minimal-surface-Neumann-hard}
    \end{align}
    where the numerical coefficients are introduced to enforce the boundary conditions with good accuracy.
    As explained above, we express $f_\text{NN}(\theta,\phi)$ and $\theta_0(\phi)$ by neural networks, and optimize them by the PINN calculation using the loss function~\eqref{loss_minimal-surface-Neumann-hard}.

    In each training epoch, the loss function is evaluated on randomly generated
collocation points in the $(\theta,\phi)$ domain. 
For the PDE residual, we first sample $\phi$ uniformly in the interval $[0,\pi/2]$.
For each chosen $\phi$, the corresponding lower edge $\theta_{0}(\phi)$ is
computed using the network $\Theta_{0}(\phi)$.
Then we sample a uniform random variable $u \in [0,1]$ and set
\begin{align}
 \theta = \theta_{0}(\phi) + u \,\Bigl(\tfrac{\pi}{2} - \theta_{0}(\phi)\Bigr).
\end{align}
This guarantees that $\theta$ is uniformly distributed in the vertical strip
between the moving boundary $\theta=\theta_{0}(\phi)$ and the AdS boundary
$\theta=\pi/2$.  The PDE residual loss is computed as the mean squared
value of the residual evaluated at these $(\theta,\phi)$ points.
The number of such interior samples per epoch is denoted $N_{\text{int}}$.
The sampling points for the Neumann boundary condition at $\phi=0,\pi/2$ are taken similarly.
   
    The PINN calculation proceeds as follows. First, we conduct a short training to make $\theta_0(\phi)$ take some value in $[0, \pi/2]$ to facilitate the following training process. In our numerical code, we choose the initial profile as $\theta_0(\phi) \sim \arctan\frac{R_\text{bdy}(\phi)}{r_0}$, 
    which is the value of $\theta_0$ if the surface were given by a cylinder that has radius $R(\phi) = R_\text{bdy}(\phi)$ and is homogeneous in the $r$ direction.
    Next, we conduct a training for $f_\text{NN}(\theta,\phi)$ and $\theta_0(\phi)$ to minimize the loss function~\eqref{loss_minimal-surface-Neumann-hard}.

\begin{figure}[htbp]
        \centering
        \subfigure[$R(\theta,\phi)$]{\includegraphics[width=7cm]{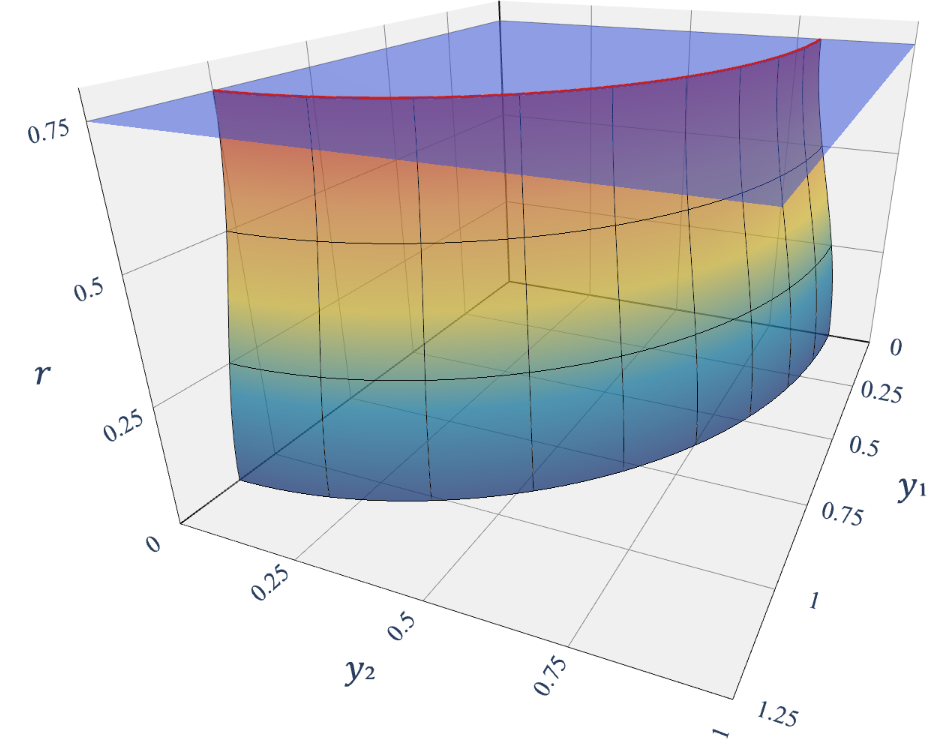}}
        \subfigure[$\theta_0(\phi)$]{\includegraphics[width=7cm]{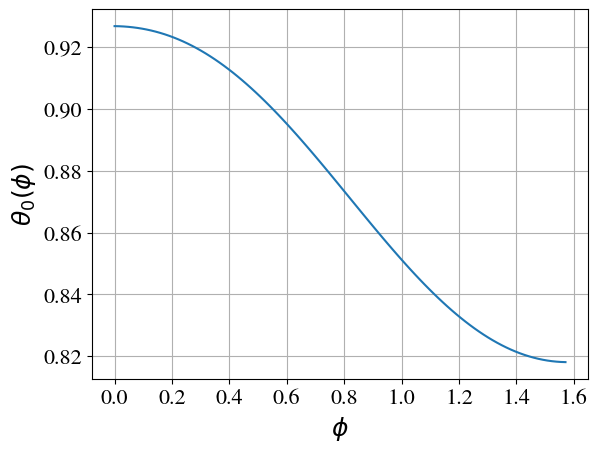}}
        \caption{Profile of $R(\theta,\phi)$ and $\theta_0(\phi)$ for $R_\text{bdy}(\phi) = 1+0.1\times \cos(2\phi)$ when the additional boundary is located at $r=0.75$ (blue plane) obtained by ``hard-enforcing'' the boundary conditions.}
        \label{fig:Neumann-hard}
    \end{figure}

    In figure~\ref{fig:Neumann-hard}, we show the numerical result of $R(\theta,\phi)$ obtained by the ``hard-enforcing" scheme described above.
    This numerical result is consistent with figure~\ref{fig:Dome-Neumann}, while we find some discrepancies. For example, the position $y_1$ of the surface at $(y_2,r) = (0,r_0 = 0.75)$ appears to differ by $\sim 4\%$.
    When the surface is axisymmetric ($R_\text{bdy} = \text{const.}$), we confirmed that the methods give the same result.
    One possible cause of this discrepancy would be the learning bias caused by the structure of the solution ansatz~\eqref{R_PINN-ansatz_Neumann-hard}. We defer a careful inspection of this subtlety to future works.


\begin{figure}[htbp]
    \centering
    \includegraphics[width=7cm]{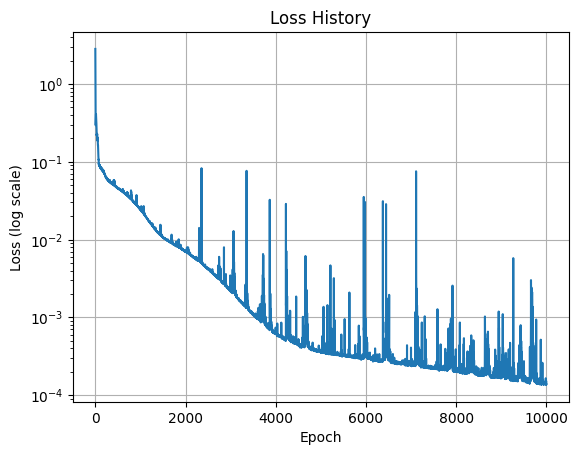}
    \caption{Loss function history for the two-dimensional minimal surface in Euclidean AdS$_3$ spacetime intersecting with the Neumann brane at $r=0.75$.}
    \label{fig:loss_round-N}
\end{figure}

\begin{table}[htbp]
  \centering
  \setlength{\tabcolsep}{7pt}
  \renewcommand{\arraystretch}{1.12}
  \begin{tabular}{lcc}
    \hline
    \textbf{Loss component} & \textbf{Value} & \textbf{Share of Total} \\
    \hline
    Total Loss                  & $2.65\times 10^{-4}$ & $100.00\%$ \\
    PDE Loss                    & $2.46\times 10^{-4}$ & $92.83\%$ \\
    BC ($f_\text{NN}$, Neumann)$\times 10$  & $5.81\times 10^{-6}$ & $2.19\%$ \\
    BC ($\theta_0(\phi)$, Neumann)$\times 10$            & $1.32\times 10^{-5}$     & $4.98\%$ \\
    \hline
  \end{tabular}
  \caption{The breakdown of the loss function~\eqref{loss_minimal-surface-Neumann-hard} at the end of the training (epoch $10^4$). The values of the loss components include the numerical coefficients introduced in \eqref{loss_minimal-surface-Neumann-hard}.}
  \label{tab:loss_round-N}
\end{table}

Figure~\ref{fig:loss_round-N} shows the loss history in the ``hard enforcing'' case. The total Loss collapses from $2.88$ to $2.65\times10^{-4}$ in $10^4$ epochs, and it is almost dominated by the PDE Loss ($2.46\times10^{-4}$ at the end of the training).
See table~\ref{tab:loss_round-N} for details.
The boundary conditions terms are maintained at the $10^{-5}$ level, indicating that the soft constraints $\partial_\phi f_\text{NN}|_{\phi=0,\ \pi/4}$ and $\partial_\phi \theta_0|_{\phi=0,\pi/4}$ are enforced to good accuracy.


\section{Minimal surface with light-like boundary}
\label{sec:light-like_loop}

In this section, we apply the construction method for minimal surfaces developed in the previous sections to a particular novel problem in high-energy physics and string theory studied in our companion paper \cite{Hashimoto2025instanton}.

We consider a minimal surface in the five-dimensional AdS spacetime that ends on a curve $C = \{\Delta y^\mu = 2\pi p_i^\mu\}$ ($i=1,\ldots,n$), where $p_i^\mu$ are light-like vectors satisfying $g_{\mu\nu} p_i^\mu p_i^\nu = 0$ and $\sum_i p_i^\mu = 0$. In the original physics problem, they correspond to the momenta of the scattering particles (gluons), and the area of the minimal surface corresponds to the logarithm of the scattering amplitude for those particles, according to \cite{Alday:2007hr}.

The difficulties of this problem are i) the minimal surface in this problem is a two-dimensional surface in five-dimensional spacetime, where the value of the time coordinate is not constant on the surface, while it was constant in the previous sections, and ii) the boundary of the minimal surface is given by a polygon with non-smooth corners.
We report our attempt to resolve these issues in this section.

\subsection{Action and equation of motion}

We consider the minimal surface in the five-dimensional AdS spacetime (AdS$_5$), whose metric is given by
\begin{equation}
ds^2 = \frac{R^2}{r^2} \left(
dy^\mu dy_\mu + dr^2
\right)
\qquad (\mu=0,1,2,3)\,,
\label{AdS5}
\end{equation}
where $y_\mu := \eta_{\mu\nu} y^\nu$, and $\eta_{\mu\nu}$ is a four-dimensional Minkowski metric.
$y_0$ is the time coordinate, and $r, y_i$ ($i=1,2,3$) are the space coordinates of this spacetime.\footnote{More precisely, $\{y^\mu\}$ are the T-dual counterpart of the original spacetime coordinates $\{x^\mu\}$, while $r$ is an inverse of the original AdS depth $z$ as $r:=R^2/z$. See \cite{Alday:2007hr} for more details.} 

We assume that the $p_i^\mu$ vectors and the minimal surfaces are all situated on the $y_3 = 0$ surface. Then, we may parameterize the minimal surface position as $y_\mu = y_\mu(\sigma^1, \sigma^2)$\, ($\mu=0,1,2$) $r=r(\sigma^1, \sigma^2)$, and $y_3 = 0$.
Here $\sigma_i$ ($i=1,2$) are the coordinates along the minimal surface called the world-sheet coordinates, which will be specified later.
In this parametrization, the induced metric on the worldsheet is given by
\begin{equation}
ds^2
=
g_{\mu\nu}
\partial_i y^\mu \partial_j y^\nu 
d\sigma^i d\sigma^j
=
\frac{R^2}{r^2}
\left(
    \eta_{\mu\nu} \partial_i y^\mu \partial_j y^\nu 
    + \partial_i r \partial_j r 
\right) d\sigma^i d\sigma^j\,,
\end{equation}
where $\partial_i := \partial / \partial \sigma^i$.
Then, the Nambu-Goto action for the minimal surface is given by
\begin{align}
S &= \int d\sigma^1 d\sigma^2
\sqrt{-\det \bigl( g_{\mu\nu}\partial_i y^\mu \partial_j y^\nu \bigr)}
\notag \\
&=
\int d\sigma^1 d\sigma^2
\frac{R^2}{r^2}\sqrt{
    -\frac12 \epsilon^{ik}\epsilon^{jl}
    \left( y_{\mu,i} y^{\mu}{}_{,j} + r_{,i} r_{,j} \right)
    \left( y_{\mu,k} y^{\mu}{}_{,l} + r_{,k} r_{,l} \right)
}
\notag \\
&
=:
\int d\sigma^1 d\sigma^2 \frac{R^2}{r^2} \sqrt{-\det h}
=:
R^2 \int d\sigma^1 d\sigma^2 L \,,
\label{L_GC}
\end{align}
where 
$h_{ij}:=  \eta_{\mu\nu} \partial_i y^\mu \partial_j y^\nu 
    + \partial_i r \partial_j r $
is the induced metric on the surface defined without the warp factor $r^{-2}$.
The tensor $\epsilon^{ij}$ is a totally anti-symmetric tensor in two dimensions with $\epsilon^{12} = 1$, and $f_{,i} := \partial_i f$.

The Euler-Lagrange equations for the Lagrangian $L$ in \eqref{L_GC} are given by
\begin{align}
    \frac{\delta L }{\delta y_\mu}
    &=
    - \partial_i \left[
        \frac{1}{2r^2 \sqrt{-\det h}}
        \frac{\partial}{\partial y_{\mu,i} }\left(-\det h\right)
    \right]
    =
    \epsilon^{ik}\epsilon^{jl} \partial_i
    \left[
        \frac{y^\mu{}_{,j} \left(
            y_{\nu,k} y^\nu{}_{,l} + r_{k} r_{,l}
        \right)}{r^2 \sqrt{-\det h}}
            \right]
    =0\,,
    \label{EoM_y_surface}
    \\
    \frac{\delta L }{\delta r}
    &=
    \frac{\partial L }{\partial r}
    - \partial_i \left(
        \frac{\partial L }{\partial r_{,i}}
    \right)
    =
    - \frac{2}{r^3} \sqrt{-\det h}
    - \partial_i \left[
        \frac{1}{2r^2 \sqrt{-\det h}}
        \frac{\partial}{\partial r_{,i} }\left(-\det h\right)
    \right]
    \notag \\
    &=
    - \frac{2}{r^3} \sqrt{-\det h}
    +\epsilon^{ik}\epsilon^{jl} \partial_i
    \left[
        \frac{r_{,j}
        \left(
            y_{\nu,k} y^\nu{}_{,l} + r_{k} r_{,l}
        \right)}{r^2 \sqrt{-\det h}}
    \right]
    =0\,.
    \label{EoM_r_surface}
\end{align}
Due to the coordinate invariance with respect to $\sigma^i$, only two equations among the above equations (for $\mu = 0,1,2$ and $r$) are independent.

By specifying the coordinates $\sigma^i$ as $\theta,\phi$ as we did in section~\ref{sec:minimal-surface} and in \eqref{sph-coords_R}, the above equations reduce to a set of two second-order elliptic PDEs for $R(\theta,\phi)$ and $y_0(\theta,\phi)$. We will formulate our numerical scheme in these coordinates.
Alternatively, we could choose the Cartesian coordinates $y_1, y_2$ as the independent variables. Some of the results below, including the exact solution \eqref{exact-soln}, are given in these coordinates.

\subsection{Boundary conditions}
\label{sec:BC_square}

We focus on the case where the minimal surface is edged by four light-like vectors on the AdS boundary. This problem is related to finding the scattering amplitude of the two-to-two scattering of gluons in the original work~\cite{Alday:2007hr}.

\subsubsection{AdS boundary}
\label{sec:BC-AdSbdy_square}

\begin{figure}[htbp]
    \centering
    \includegraphics[width=8cm]{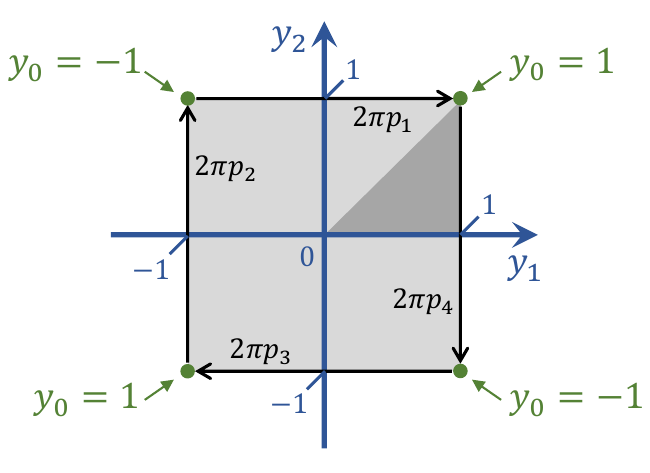}
    \caption{Boundary condition on the AdS boundary.
    The minimal surface fill the region $(y_1, y_2) \in [-1,1]\times [-1,1]$ (light gray). Along the boundary at $y_{1,2}=\pm 1$, $r$ is set to zero and $y_0$ is specified as \eqref{eq:bc_rhombus}.
    We take the region $0 \leq y_1 \leq 1 \,  \cap \, y_2\leq y_1$ (dark gray) as the numerical domain based on the symmetry of the problem.    }
    \label{fig:BC}
\end{figure}

We consider the simplest case where the edge of the minimal surface is on a loop of four light-like vectors $C = \{\Delta y^\mu = 2\pi p_i^\mu\}$ on the AdS boundary, where the four vectors are in $(y_0, y_1, y_2)$ space and given by
\begin{equation}
    2\pi p_1 = (2,2,0)\,, \quad
    2\pi p_2 = (-2,0,2)\,, \quad
    2\pi p_3 = (2,-2,0)\,, \quad
    2\pi p_4 = (-2,0,-2)\,.
\end{equation}
In terms of the dynamical variables $r$ and $y_0$, we impose the Dirichlet conditions at the edge of the minimal surface as
\begin{equation}
    y^\mu \in  \bigl\{\Delta y^\mu = 2\pi p_i^\mu\,, i=1,\ldots, 4 \bigr\}
    \,,
    \qquad
    r = 0\,.
\end{equation}
More explicitly, we assume that $r$ and $y_0$ on the edge of the minimal surface are given by
\begin{equation}
    r(y_1 = \pm 1, y_2) = r(y_1, y_2 = \pm 1) = 0\,,
    \quad
    y_0(y_1 = \pm 1, y_2) = \pm y_2\,,
    \quad
    y_0(y_1, y_2 = \pm 1) = \pm y_1\,.
    \label{eq:bc_rhombus}
\end{equation}
See figure~\ref{fig:BC}.

\subsection{Symmetries and regularity}
\label{sec:BC-symmetry_square}

The problem setting has the following symmetries:
\begin{itemize}
    \item reflection symmetry with respect to $y_1 = \pm y_2$
    \item reflection symmetry with respect to $y_1 = 0, y_2 = 0$ with a sign flip $y_0 \to -y_0$.
\end{itemize}
These symmetries and the boundary conditions at the AdS boundary imply that $r(y_1, y_2)$ and $y_0(y_1, y_2)$ enjoy the same symmetries.

Thanks to the symmetry, it suffices to solve the problem only on an eighth of the whole coordinate domain. Below, we choose to solve the problem in the triangular region defined by $y_1, y_2 \geq 0$ and $y_1 \geq y_2$.
On the borders given by the $y_1$ axis and the $y_1 = y_2$ line, based on the symmetry of the problem, $r(y_1, y_2)$ should behave as an even function in the perpendicular direction to them.
As for $y_0(y_1,y_2)$, it is an odd function with respect to the $y_1$ axis, while it is an even function with respect to the $y_1 = y_2$ line.

We also need to ensure that the surface is smooth at the origin $(y_1, y_2) = 0$, particularly when we solve the problem only in the region defined above.
Both $y_0(y_1, y_2)$ and $r(y_1,y_2)$ should satisfy the Neumann boundary conditions in the radial direction from the origin to maintain the regularity.

\subsubsection{Exact solution}

Under the above boundary conditions, the equations of motion \eqref{EoM_y_surface}, \eqref{EoM_r_surface} admit an exact solution given by 
\begin{equation}
    y_0 = y_1 y_2, \qquad r=\sqrt{(1-y_1{}^2) (1-y_2{}^2) }\,.
    \label{exact-soln}
\end{equation}
 The derivation of this solution is given in \cite{Alday:2007hr}. See figure~\ref{fig:exact_solution} in appendix~\ref{app:square-alt} for its shape in the background spacetime.
We will compare our numerical solutions with it and also employ it as a building block in the numerical scheme introduced in the next subsection.

\subsubsection{Neumann boundary conditions at $r=r_0$}
\label{sec:BC-Neumann_square}

As an extension of the problem to construct minimal surfaces satisfying the above Dirichlet condition, we also consider the case where the minimal surface intersects a flat boundary at $r=r_0$.
This case corresponds to the insertion of an instanton in the gluon scattering amplitude, as argued in our companion paper~\cite{Hashimoto2025instanton}.

It can be shown that the surface must intersect the boundary orthogonally to maintain the minimality of the surface area.
In terms of the variables used here, this boundary condition is expressed as
\begin{equation}
    R(\theta,\phi) \cos\theta = r_0\,, \qquad
    \bigl(R(\theta,\phi) \sin\theta \bigr)_{\,\theta} = 0\,, \qquad
    y_0{}_{,\theta} = 0
\end{equation}
on the boundary at $r=r_0$.

To facilitate the numerical construction for surfaces satisfying these conditions, we define a function $\theta_0(\phi)$ by
\begin{equation}
    R\bigl(\theta = \theta_0(\phi),\phi\bigr) \cos\theta_0(\phi) = r_0\,.
\end{equation}
By definition, 
$\theta_0(\phi)$ depends on the function form of $R(\theta,\phi)$.
Then, the problem to construct a minimal surface in this case becomes a boundary value problem to solve the PDEs \eqref{EoM_y_surface}, \eqref{EoM_r_surface} for $R(\theta,\phi), y_0(\theta,\phi)$ over the domain $(\theta,\phi) \in [\theta_0(\phi),\pi/2] \times [0,2\pi]$.

\subsection{Numerical method}

\subsubsection{Solution ansatz}
\label{sec:ansatz_square}

One of the difficulties with the problem in this section is that the edge of the surface has non-smooth corners, whereas the surface is smooth everywhere away from the edge. To accommodate the non-smoothness at the edge, we construct baseline and envelope functions tailored to the edge shape.
By means of the series expansion near the boundaries and the coordinate transform, we find that the general solution may be expressed as
\begin{align}
    R(\theta,\phi) &= \mathcal{B}_{R}(\theta,\phi) + \mathcal{E}_R(\theta,\phi) \times  f_r(\theta,\phi)\, ,
    \label{R-ansatz_square}
    \\
        y_0(\theta,\phi) &=
    \mathcal{B}_{y_0}(\theta,\phi)
    + \mathcal{E}_{y_0; y_0}(\theta,\phi)\times f_{y_0}(\theta,\phi) 
    + \mathcal{E}_{y_0; r}(\theta,\phi) \times f_{r}(\theta,\phi)
    \,,
    \label{y0-ansatz_square}
\end{align}
where
\begin{gather}
    \mathcal{B}_{R} \equiv \sqrt{\frac{2}{1 +\rho}}\,,
    \qquad
    \mathcal{B}_{y_0} \equiv \frac{\sin^2\theta \sin(2\pi)}{1 + \rho}\,,
    \\
    \mathcal{E}_R \equiv
    \frac{2\sqrt{2}\cos^2\theta\left(1 + \frac{\cos^2\theta}{\rho}\right) }{(1 + \rho)^{3/2}}
    \,,
    \quad
    \mathcal{E}_{y_0; y_0} \equiv \frac{2 \cos^2 \theta}{1+\rho}\,,
    \quad
    \mathcal{E}_{y_0; r} \equiv \frac{2 \sin^2\theta \sin(2\phi)}{\sqrt{2+2\rho}} \times \mathcal{E}_R(\theta,\phi)\,,
\end{gather}
for which we defined
\begin{equation}
    \rho \equiv \sqrt{1 - \sin^4 \theta \sin^2(2\phi)}\,.
\end{equation}
See Appendix~\ref{App:ansatz} for the derivation.

In our numerical code, we express $f_r$ and $f_{y_0}$ in \eqref{R-ansatz_square} and \eqref{y0-ansatz_square} by neural networks, and solve the Euler-Lagrange equations \eqref{EoM_y_surface} and \eqref{EoM_r_surface} by PINN.

\subsubsection{Standard minimal surface}

We first consider a standard minimal surface that covers the whole domain of $|y_1|, |y_2| \leq 1$ as a test of our numerical scheme.\footnote{See~\cite{Dobashi:2008ia,Dobashi:2009sj} for earlier numerical studies on this type of surface.}
We solve the problem only in the domain explained in section~\ref{sec:BC-symmetry_square}, which corresponds to the region with $0\leq \theta\leq\pi/2$ and $0\leq \phi \leq \pi/4$.

In this setting, we need to impose the following boundary conditions.
\begin{itemize}
    \item Neumann boundary conditions for $R(\theta,\phi)$ at $\phi=0, \pi/4$ and $y_0(\theta,\phi)$ at $\phi=\pi/4$:
    \begin{equation}
        \partial_\phi R(\theta,\phi = 0,\pi/4) = \partial_\phi y_0(\theta,\phi=\pi/4) = 0
    \end{equation}
    \item Dirichlet boundary condition for $y_0$ at $\phi=0$:
    \begin{equation}
        y_0(\theta, \phi = 0) = 0
    \end{equation}
    \item Regularity at the center:
    \begin{equation}
        \partial_\theta y_0(\theta=0,\phi) = \partial_\theta r(\theta=0,\phi) = 0\,, \quad
        \partial_\phi y_0(\theta=0,\phi) = \partial_\phi r(\theta=0,\phi) = 0
    \end{equation}
    The second set of conditions is necessary to ensure that $y_0$ and $R$ are single-valued functions at $\theta=0$.
\end{itemize}
As for the boundary condition at the AdS boundary studied in section~\ref{sec:BC-AdSbdy_square}, we do not need to impose any condition on $f_r, f_{y_0}$ because they are satisfied by the construction of the solution ansatz \eqref{R-ansatz_square}, \eqref{y0-ansatz_square}.

In our numerical method, we impose the regularity conditions at $\theta=0$ in a hard manner, and impose the boundary conditions at the borders $\phi=0,\pi/4$. More explicitly, we express $f_r, f_{y_0}$ as
\begin{equation}
    f_r(\theta, \phi) = c_r + \theta^2 \tilde f_r(\theta,\phi)\,,
    \qquad
    f_{y_0}(\theta,\phi) = \theta^2 \tilde f_{y_0}(\theta,\phi)\,,
\end{equation}
where $c_r$ is a constant.
The regularity conditions at $\theta=0$ are automatically satisfied as long as $c_r, \tilde f_r, \tilde f_{y_0}$ are finite there.
We express $\tilde f_r$ and $\tilde f_{y_0}$ by neural networks, and optimize $\tilde f_r$, $\tilde f_{y_0}$, and $c_r$ by PINN with the following loss function:
\begin{align}
\text{Loss}
&=
\quad
\frac{1}{N_\text{int}}\sum_{\Theta_{10^{-3}}\times\Phi}
\bigl[\tanh^2(\text{PDE}_{R})+\tanh^2(\text{PDE}_{y_0})\bigr]
\notag \\
&\quad
+
\frac{1}{N_{bc}}\sum_{\Theta_0}
\Bigl[
(\partial_\phi \tilde f_r(\theta,0))^2 + (\partial_\phi \tilde f_r(\theta,\tfrac{\pi}{4}))^2
+ (\partial_\phi \tilde f_y(\theta,\tfrac{\pi}{4}))^2
+ c_D\,\tilde f_y(\theta,0)^2
\Bigr]\,,
\label{loss_square}
\end{align}
where \(\Theta_{\varepsilon}\equiv(\varepsilon,\tfrac{\pi}{2}-\varepsilon)\),  $\Phi = (0,\pi/4)$ and \(c_D=10\).
We introduce a small cutoff $\varepsilon = 10^{-3}$ near $\theta=0,\pi/2$ for the PDE residuals, while we use the full range with $\varepsilon = 0$ for the boundary terms.
The summations in the loss function are evaluated 
for \(N_{\text{int}}=4000\) interior samples \((\theta_i,\phi_i)\) and \(N_{\text{bc}}=256\) \(\theta\)-samples, which are chosen randomly and uniformly over $\Theta_{10^{-3}} \times \Phi$ and $\Theta_0$, respectively.
For the PDE loss term in the total loss function, the $\tanh$ envelope was introduced to bound individual interior contributions by \(1\), mitigating outliers while remaining quadratic near zero.

\begin{figure}[htbp]
    \centering
    \subfigure[$y_0$]{\includegraphics[width=7cm]{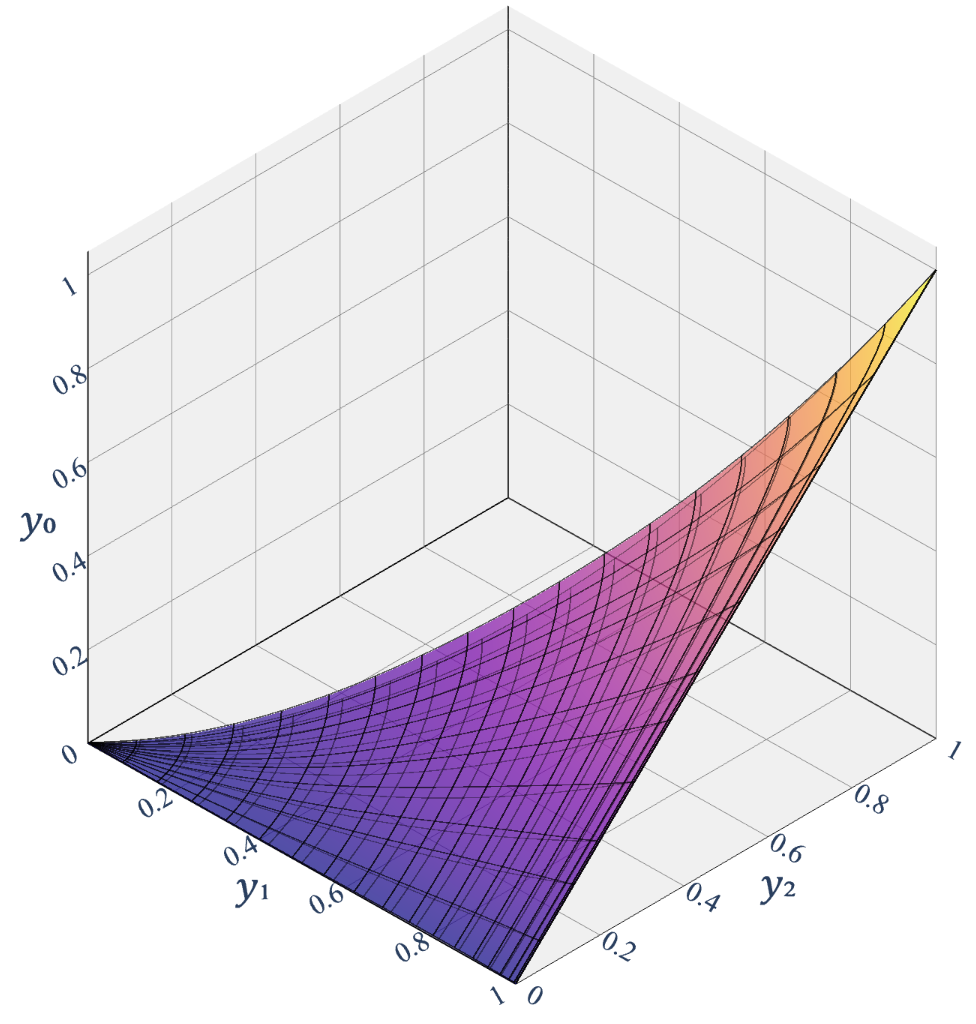}}
    \quad
    \subfigure[$R$]{\includegraphics[width=7cm]{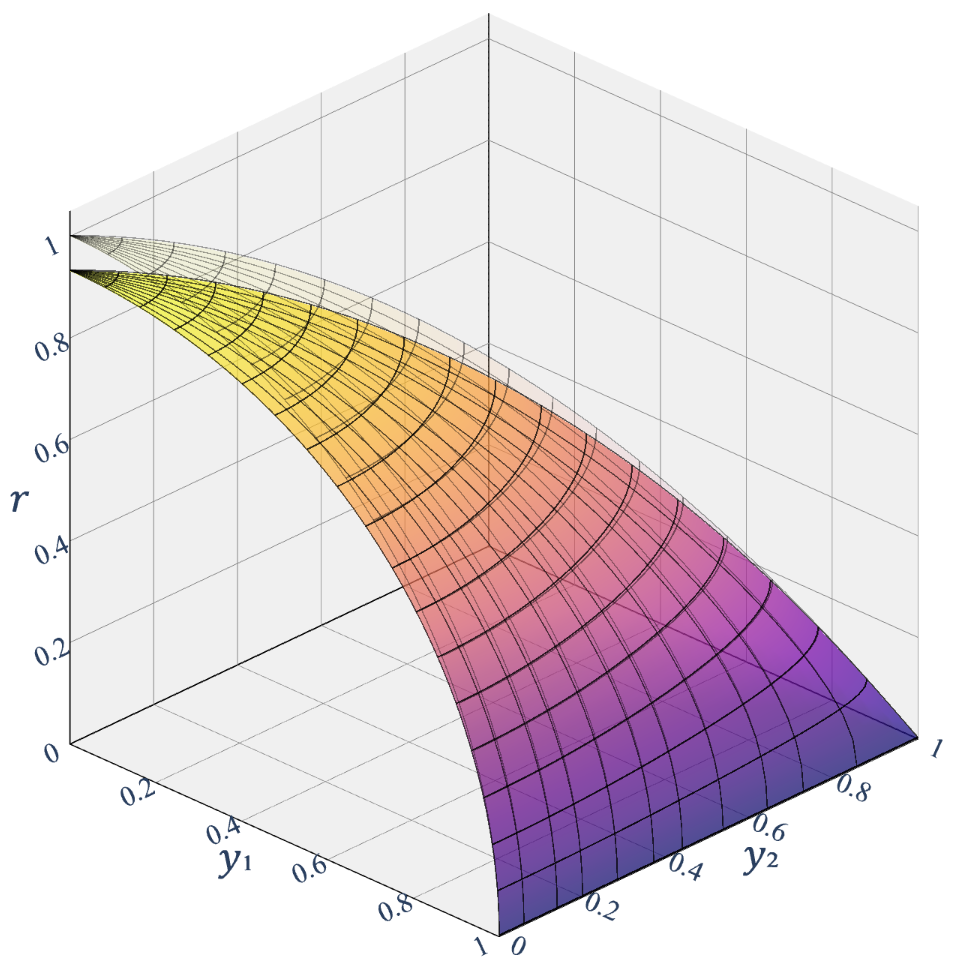}}
    \caption{Plots of $y_0$ and $r$ for the standard minimal surface spanned by the light-like loop. Only an eighth of the surface ($0\leq\phi\leq \pi/4$) is shown. The exact solution~\eqref{exact-soln} is overlaid as transparent grid lines. The grid lines correspond to constant $\theta$ and $\phi$.}
    \label{fig:y0-R_square}
\end{figure}

In figure~\ref{fig:y0-R_square}, we show the numerical results obtained by PINN.
The exact solution \eqref{exact-soln} will be reproduced if $f_{y_0}$ and $f_r$ vanish, while they develop values smaller than $O(1)$ in the numerical solutions. The peak height of $r$ (i.e., the value of $r(\theta=0,\phi)$) is unity for the exact solution, while it is $r(\theta=0,\phi) =0.932$; the relative error of the $r$ surface shape is estimated as $\sim 7\%$ at least with respect to its height.

\begin{figure}[htbp]
    \centering
    \includegraphics[width=10cm]{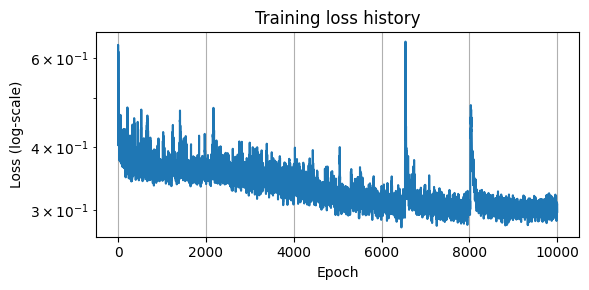}
    \caption{Loss function history for the standard minimal surface spanned by the light-like loop.}
    \label{fig:loss_square}
\end{figure}

\begin{table}[t]
  \centering
  \setlength{\tabcolsep}{8pt}
  \renewcommand{\arraystretch}{1.12}
  \begin{tabular}{lcc}
    \hline
    \textbf{Loss component} & \textbf{Value (epoch $10^4$)} & \textbf{Share of Total} \\
    \hline
    Total                        & $2.98\times 10^{-1}$ & $100\%$ \\
    PDE($y_0$)                         & $2.81\times 10^{-1}$ & $94.4\%$ \\
    PDE($R$)                         & $1.63\times 10^{-2}$ & $5.47\%$ \\
    BC $(f_{r})$                 & $2.36\times 10^{-4}$ & $0.079\%$ \\
    BC $(\tilde f_{y_0}\text{, Neumann})$ & $3.31\times 10^{-4}$ & $0.111\%$ \\
    BC $(\tilde f_{y_0}\text{, Dirichlet})$ & $3.59\times 10^{-5}$ & $0.012\%$ \\
    \hline
  \end{tabular}
  \caption{Breakdown of the loss~\eqref{loss_square} at the end of training (epoch $10^4$).
  Percent shares are computed relative to \emph{Total}.
  The two $\tilde f_{y_0}$ entries correspond to the Neumann (N) and Dirichlet (D)
  parts of the boundary constraints. The learned offset parameter at this epoch is
  $c_R\simeq -3.42\times 10^{-2}$.}
  \label{table:loss_square}
\end{table}

Table~\ref{table:loss_square} shows the breakdown of the total loss~\eqref{loss_square} at the end of the learning (epoch $10^4$).
The total loss is dominated by the PDE residuals; the residual for the Euler-Lagrange equation for $y_0$ dominates $\sim$94\% of the total, while that for $R$ contributes about $5.5\%$.
Relative to initialization (epoch 0), the total loss decreased from
$5.00\times 10^{-1}$ to $2.98\times 10^{-1}$ ($\sim40\%$ reduction); the PDE loss for $R$
dropped by nearly an order of magnitude ($1.45\times 10^{-1}\!\to\!1.63\times 10^{-2}$),
whereas that for $y_0$ decreased more modestly ($3.48\times 10^{-1}\!\to\!2.81\times 10^{-1}$).
These features change slightly by changing the numerical coefficients in the total loss function, but the general tendency and the numerical solution quality do not change qualitatively. This fact may indicate that more improvements at the level of formulation (such as solution ansatz and coordinate choices) are necessary to realize better numerical results.

As for the boundary-condition penalties for $f_r$ and $\tilde f_{y_0}$ are each $\ll 1\%$
of the total (combined $\lesssim 0.2\%$), indicating that the Dirichlet/Neumann
constraints are effectively saturated at the end of training.

\subsection{Introducing Neumann boundary}

We report on our attempt to generate solutions when we introduce the Neumann boundary, for which we need to impose the boundary conditions given in section~\ref{sec:BC-Neumann_square} at the moving boundary $\theta=\theta_0(\phi)$.
This problem is of particular interest, because in our companion paper \cite{Hashimoto2025instanton} we need to solve this situation for evaluating a physics problem: an instanton effect in gluon scattering amplitudes.
To solve this problem, we use a method similar to the ``hard enforcing'' method used in section~\ref{sec:minimal-surface_hard}.\footnote{Approaches based on the ``soft enforcing'' method in section~\ref{sec:minimal-surface_soft} typically suffered from strong numerical errors and produced unphysical solutions, although there may be some way to improve them.}
In appendix~\ref{app:square-alt}, we report numerical results based on an approach different from this section and discuss the choice of the PINN architecture.

\subsubsection{Numerical domain with moving boundary}
We work in $(\theta,\phi)$ with
\begin{equation}
  \theta \in \left[\theta_0(\phi),\frac{\pi}{2}-\varepsilon_\theta\right],\qquad
  \phi \in \left[0,\frac{\pi}{4}\right],\qquad
  \varepsilon_\theta=10^{-3}\ ,
\end{equation}
where $\varepsilon_\theta$ is a cutoff for the AdS boundary. The moving boundary $\theta_0(\phi)$ is learned as
\begin{equation}
  \theta_0(\phi)
  = B_{\theta_0}(\phi)+\phi^2\left(\frac{\pi}{4}-\phi\right)^2\,f_{\theta_0}(\phi),
\end{equation}
where
\begin{equation}
B_{\theta_0}(\phi) \equiv  
\theta_1
+\frac{16 (3 \pi -8 \phi ) \phi ^2}{\pi ^3}\Delta\theta_1
\end{equation}
is a cubic polynomial
that satisfies 
$B_{\theta_0}(\phi=0) = \theta_1$, $B_{\theta_0}(\phi=0) = \theta_1 + \Delta \theta_1$,
and the Neumann boundary conditions at $\phi = 0,\pi/4$.
We express $f_{\theta_0}$ by a small neural network, and optimize $f_{\theta_0}$ and also the constants $\theta_1$ and $\Delta \theta_1$ by training.

\subsubsection{Solution ansatz and neural network}

We use the solution ansatz \eqref{R-ansatz_square}, \eqref{R-ansatz_square} again, and impose the boundary conditions explained in section~\ref{sec:BC-Neumann_square}.
We impose most of the boundary conditions in the ``soft'' sense by the penalty terms in the total loss function.
Only for the Dirichlet condition at $\phi=0$ on $y_0$, we enforce it in the ``hard'' sense by re-defining $f_{y_0}$ as
\begin{equation}
    f_{y_0} = \sin(2\phi) \tilde f_{y_0}\,.
\end{equation}
Also, based on the symmetry described in section~\ref{sec:BC_square} and the structures of the baseline and envelope functions defined in section~\ref{sec:BC-Neumann_square}, we impose 
\begin{equation}
\partial_\phi f_r(\theta,0)=\partial_\phi f_r(\theta,\tfrac{\pi}{4})=0,\qquad
\partial_\phi \tilde f_{y_0}(\theta,0)=\partial_\phi \tilde f_{y_0}(\theta,\tfrac{\pi}{4})=0. 
\end{equation}

Both $f_r$ and $\tilde f_{y_0}$
are represented by small fully connected neural networks
$\mathsf{MLP}_{r},\mathsf{MLP}_{y_0}:\mathbb{R}^2\!\to\!\mathbb{R}$ that take
$(\theta,\phi)$ as input and output a scalar. Each MLP has four hidden layers of width~64
with $\tanh$ activations. We set
\begin{equation}
f_r(\theta,\phi) = \mathrm{softplus}\big(\mathsf{MLP}_{r}(\theta,\phi)\big),
\qquad
\tilde f_{y_0}(\theta,\phi) = \mathsf{MLP}_{y_0}(\theta,\phi)\,,
\end{equation}
where we applied the $\mathrm{softplus}$ function defined as
$\mathrm{softplus}(x)=\log\bigl(1+e^{ x}\bigr)$.
The $\mathrm{softplus}$ function guarantees $f_r>0$ while preserving
smooth first and second derivatives needed by the PDE residuals; 
$\tilde f_{y_0}$
is left
unconstrained.

The restriction $f_r > 0$ is introduced to stabilize the numerical calculation, and it is consistent with the results in sections~\ref{sec:WL} and \ref{sec:minimal-surface}
that the minimal surface is always shifted toward the Neumann boundary when it is introduced. It would be desirable to prove this property mathematically, but it is beyond the scope of this work. 

\subsubsection{Loss function}
The total loss is a weighted sum of interior PDE terms, inner-wall constraints, and edge symmetries:
\begin{align}
\mathcal L 
&=
\quad
\frac{1}{N_{\mathrm{int}}}
\sum_{j=0}^{N_{\phi}^{\mathrm{in}}-1}\;
\sum_{k=0}^{N_{u}^{\mathrm{in}}-1}
\left[
 \tanh^2\big(\text{PDE}_{y_0}(\theta_{jk},\phi_j)\big)
+ 5\,\tanh^2\big(\text{PDE}_{R}(\theta_{jk},\phi_j)\big)
\right]
\notag \\
 &\quad
 +\frac{1}{N_{\phi}^{(0)}}
 \sum_{m=0}^{N_{\phi}^{(0)}-1}\left\{
20 \big(R\cos\theta-r_0\big)^2
+ \big[ \partial_\theta (R\sin\theta) \big]^{2}
+\big( \partial_\theta y_0\big)^{2}
\right\}_{\theta=\theta_0(\tilde\phi_m),\,\phi=\tilde\phi_m}
\notag \\
&\quad
 +\frac{1}{2N_{u}^{\mathrm{bc}}}
 \sum_{\phi = 0,\pi/4}\;
 \sum_{\ell=0}^{N_{u}^{\mathrm{bc}}-1}
 \left[
  \big(\partial_\phi f_r\big)\big(\theta^{\mathrm{edge}}_\ell(\phi),\phi\big)^{2}
+ \big(\partial_\phi \tilde f_{y_0}\big)\big(\theta^{\mathrm{edge}}_\ell(\phi),\phi\big)^{2}
\right]\,,
\label{loss_square-N}
\end{align}
where $N_{\mathrm{int}}=N_{\phi}^{\mathrm{in}}N_{u}^{\mathrm{in}}$.
The numerical coefficients of each term are chosen to realize stable numerical calculation and to ensure that the boundary conditions are satisfied to good accuracy.

To evaluate the PDE terms in the loss function~\eqref{loss_square-N}, we use the following coordinate grids:
%
\begin{equation}
\Phi^{\mathrm{in}}
=\Bigl\{\phi_j=\tfrac{j}{N_{\phi}^{\mathrm{in}}-1}\,\tfrac{\pi}{4}\Bigr\}_{j=0}^{N_{\phi}^{\mathrm{in}}-1},
\qquad
U^{\mathrm{in}}
=\Bigl\{u_k=u_{\min}+\tfrac{k}{N_{u}^{\mathrm{in}}-1}\,(u_{\max}-u_{\min})\Bigr\}_{k=0}^{N_{u}^{\mathrm{in}}-1},
\end{equation}
with $u_{\min}=10^{-6}$ and $u_{\max}=1-10^{-6}$.
The interior points are then specified as
\begin{equation}
\theta_{jk}=\theta_0(\phi_j)+\biggl(\frac{\pi}{2}-\varepsilon_\theta-\theta_0(\phi_j)\biggr) u_k, 
\qquad \varepsilon_\theta=10^{-3}.
\end{equation}
At each point, the partial derivatives of $y_0, R$ are computed by automatic differentiation.

For the terms associated with the boundary conditions at the moving boundary $\theta=\theta_0(\phi)$, we use a denser angular grid
\begin{equation}
\Phi^{(0)}=\biggl\{\tilde\phi_m=\frac{m}{N_{\phi}^{(0)}-1}\,\frac{\pi}{4}\biggr\}_{m=0}^{N_{\phi}^{(0)}-1},    
\end{equation}
and for the boundaries at $\phi=0, \pi/4$, we use
%
\begin{equation}
U^{\mathrm{bc}}
=\Bigl\{\hat u_\ell=u_{\min}+\tfrac{\ell}{N_{u}^{\mathrm{bc}}-1}\,(u_{\max}-u_{\min})\Bigr\}_{\ell=0}^{N_{u}^{\mathrm{bc}}-1},
\quad
\theta_\ell^{\mathrm{edge}}(\phi_\star)=\theta_0(\phi)
+\Bigl(\tfrac{\pi}{2}-\varepsilon_\theta-\theta_0(\phi)\Bigr)\hat u_\ell\,.
\end{equation}

We fix the sampling point numbers as $N_{\phi}^{\mathrm{in}}=64$, $N_{u}^{\mathrm{in}}=64$, 
$N_{\phi}^{(0)}=256$, and $N_{u}^{\mathrm{bc}}=256$.
Sampling is performed in the $(\phi,u)$ coordinates and then mapped to $(\theta,\phi)$
using the current moving boundary $\theta_0(\phi)$.

The $\phi$- and $u$-grids are \emph{fixed} (no random resampling), but the interior sampling points move each epoch because the mapping 
$\theta=\theta_0(\phi)+u\bigl(\frac{\pi}{2}-\varepsilon_\theta-\theta_0(\phi)\bigr)$
uses the \emph{current} $\theta_0(\phi)$. This keeps the density in $\theta$ uniform in $u$
and aligned with the evolving strip $[\theta_0(\phi),\pi/2-\varepsilon_\theta]$.

\subsubsection{Initialization}



Before the full PDE training, we perform a brief pretraining (500 steps, Adam with fixed
$\eta=10^{-3}$) on $f_r$ and $\tilde f_{y_0}$, freezing the moving boundary $\theta_0(\phi)$ with $\theta_1 \sim 0.8 $, $\Delta\theta_1 = 0$, and $f_{\theta_0}(\phi)$ is randomly initialized.
We sample the strip
$\{(\theta,\phi):\ \phi\in[0,\frac{\pi}{4}],\ \theta=\theta_0(\phi)+u(\frac{\pi}{2}-\varepsilon_\theta-\theta_0(\phi)),\ u\in(0,1)\}$
on a Cartesian grid of $(\phi,u)$ with $(N_\phi^{\mathrm{in}},N_u^{\mathrm{in}})=(64,64)$.
The targets are simple, scale-setting templates
\[
f_{r}^{\star}(\theta,\phi)=\frac{A}{\theta^2},\qquad
\tilde f_y^{\star}(\theta,\phi)=\frac{A}{\theta^2},
\]
with $A=0.1$. We minimize the mean-squared error
\begin{equation}
    \mathcal L_{\mathrm{pre}}=\big\langle\big(f_r-f_r^{\star}\big)^2\big\rangle+
\big\langle\big(f_{y_0}-f_{y_0}^{\star}\big)^2\big\rangle
\end{equation}
over the interior collocation set; no boundary losses or PDE terms are used here.
Note that $f_R$ still passes through a softplus layer, guaranteeing $f_R>0$
during pretraining. This warm start provides reasonable magnitudes and curvature for
$R$ and $y_0$,
improving the conditioning of higher-order derivatives before imposing the full PDE and
boundary constraints.

\subsubsection{Numerical results}

\begin{figure}[htbp]
    \centering
    \subfigure[$y_0(y_1, y_2)$]{\includegraphics[width=6cm]{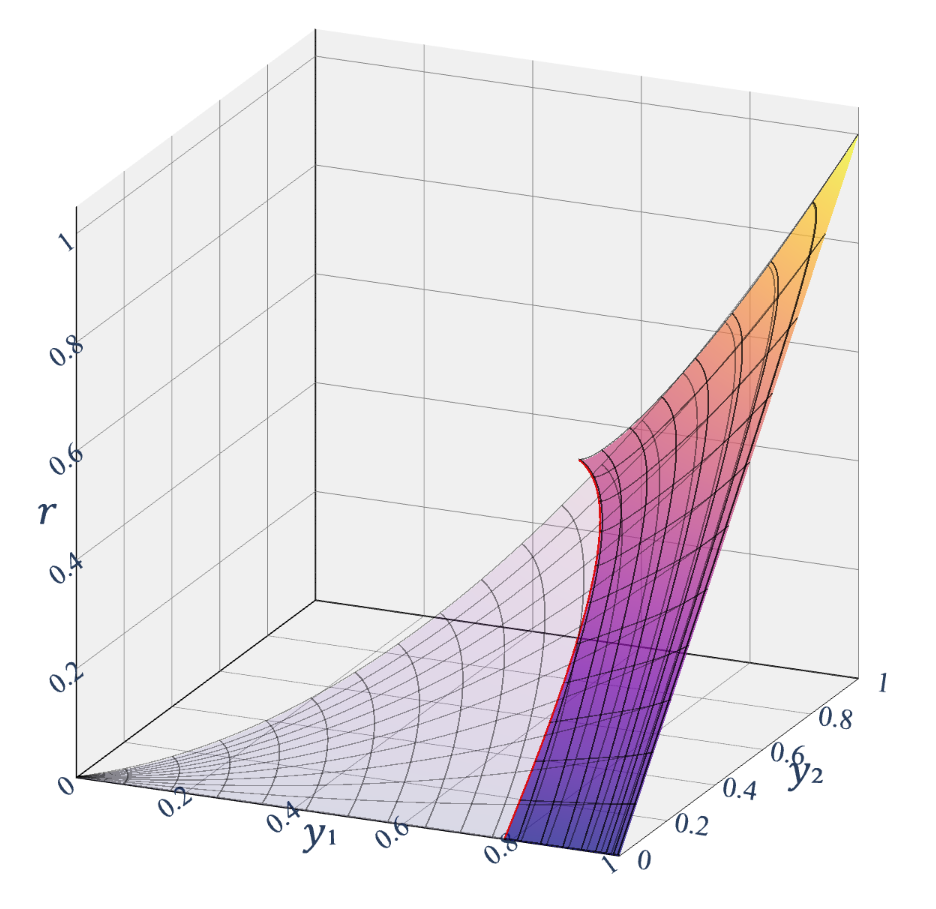}\label{fig:y0_square-N}}
    \qquad
    \subfigure[$r(y_1,y_2)$]{\includegraphics[width=6cm]{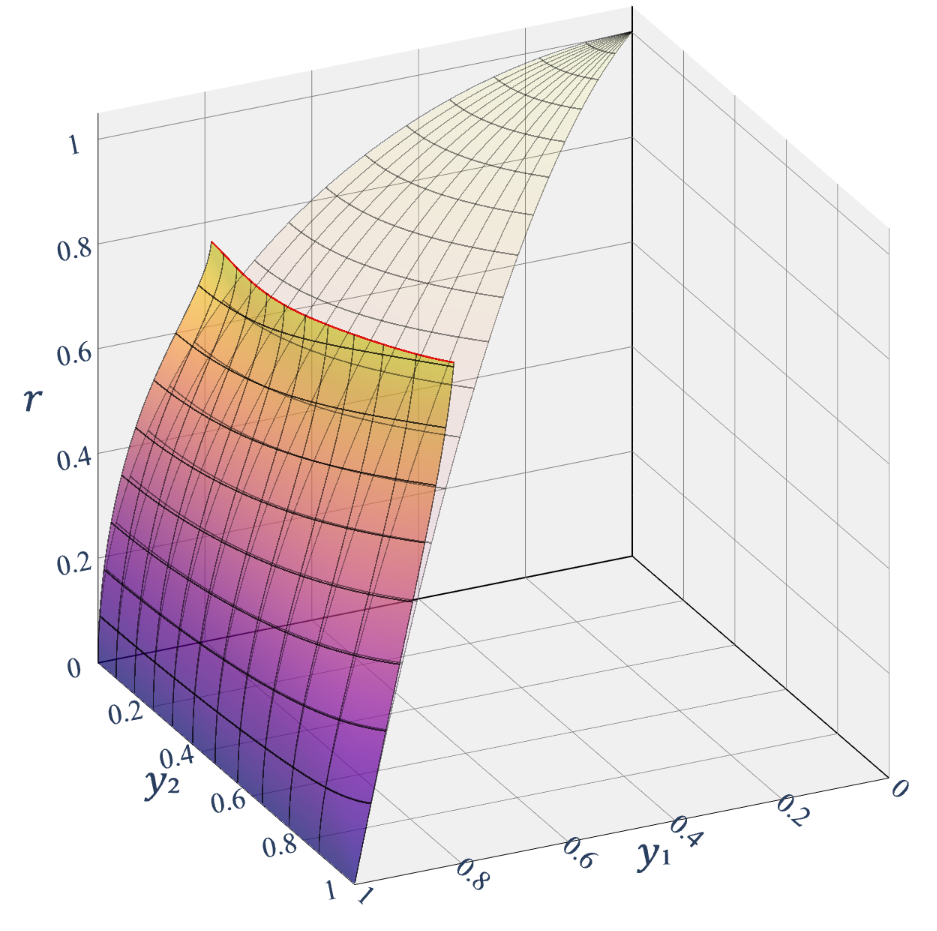}\label{fig:R_square-N}}
    \subfigure[$y_0(\phi,r)$]{\includegraphics[width=6cm]{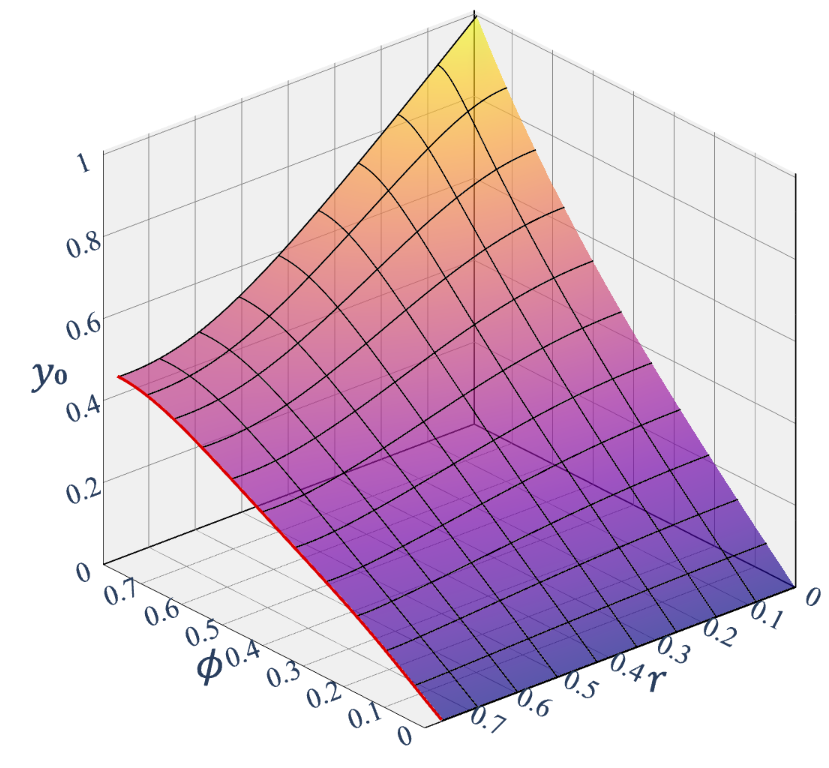}\label{fig:y0-in-r_square-N}}
    \caption{Plots of $y_0$ and $r$ for the minimal surface spanned by the light-like loop intersecting a Neumann boundary at $r=0.75$. 
    Panels (a) and (b) show $y_0$ and $r$ with respect to $(y_1, y_2)$, in which the constant-$\theta$ and -$\phi$ lines are drawn as the grid lines.
    Panel (c) shows $y_0$ with respect to $(\phi,r)$, where the grid lines are drawn on constant-$\phi$ and -$r$ lines.
    Only an eighth of the surface ($0\leq\phi\leq \pi/4$) is shown. The exact solution~\eqref{exact-soln} is overlaid as transparent grid lines in panels (a) and (b). }
    \label{fig:y0-R_square-N}
\end{figure}

Figure~\ref{fig:y0-R_square-N} presents the trained fields \(R\) and \(y_0\) over the computational domain. The surfaces are smooth across the interior and remain regular up to the moving edge \(\theta=\theta_0(\phi)\).
In figures~\ref{fig:y0_square-N} and \ref{fig:R_square-N}, we can observe that the Neumann boundary condition is well satisfied at the edge $\theta=\theta_0(\phi)$ shown by the red curve.\footnote{The obtained shape of the minimal surface is what is expected from the generic argument in the AdS/CFT correspondence, see our companion paper \cite{Hashimoto2025instanton}.}
Correspondingly, the values of the loss terms corresponding to this condition are well suppressed, as we discuss later.

\begin{figure}
    \centering
    \subfigure[$\theta_0(\phi)$]{\includegraphics[width=7cm]{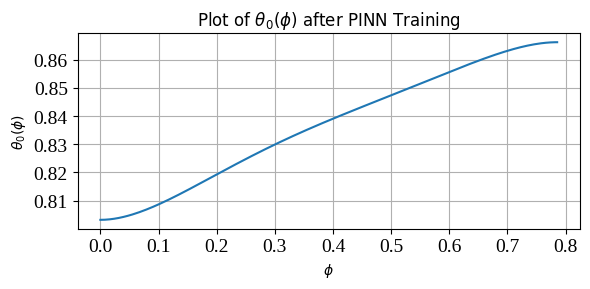}\label{fig:theta0_square-N}}
    \quad
    \subfigure[$r\bigl(\theta_0(\phi)\bigr)$]{\includegraphics[width=6.5cm]{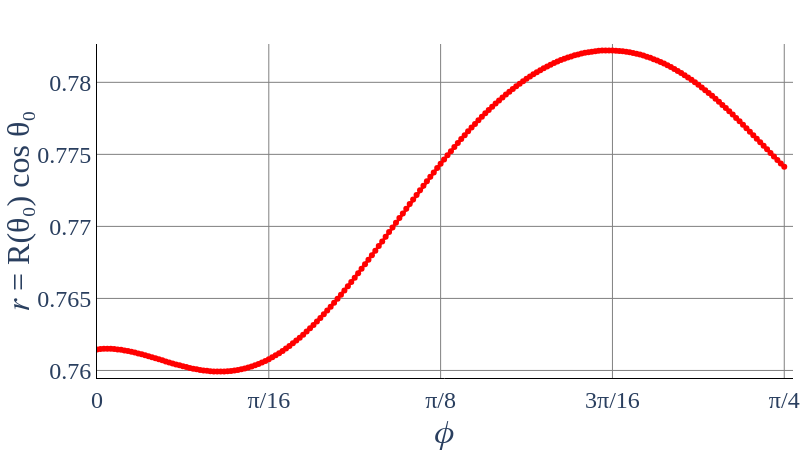}\label{fig:r-at-theta0_square-N}}
    \caption{Panel (a): profile of $\theta_0(\phi)$ after the training for minimal surface spanned by the light-like loop intersecting a Neumann boundary at $r_0=0.75$. Panel (b): profile of $r(\theta_0\bigl(\phi)\bigr)$ after the training. The largest deviation from the target value $r_0 = 0.75$ is attained at $\phi \sim 0.58$, at which $r = R(\theta_0) \cos\theta_0 \sim 0.782$, which differs from the target value by $\sim 4\%$.}
    \label{fig:theta0-and-r_square-N}
\end{figure}

The learned profile \(\theta_0(\phi)\) is shown in figure~\ref{fig:theta0_square-N}.
We can confirm that $\theta_0(\phi)$ satisfies the Neumann boundary condition at $\phi=0,\pi/4$, as we enforced it by construction. Also, $\theta_0(\phi)$ is monotonically increasing with respect to $\phi$, which is consistent with the fact that the edge on the AdS boundary $\theta=\pi/2$ is rectangular-shaped and that the moving edge at $\theta=\theta_0(\phi)$ on the surface $r$ is deformed toward that rectangular shape.
Figure~\ref{fig:r-at-theta0_square-N} shows the height of the $r$ surface at the moving edge $\theta = \theta_0(\phi)$ after the training, which should be $r = R\cos\theta = r_0 = 0.75$  if the training is perfectly accomplished. The largest deviation from the target value $r_0 = 0.75$ was $r \sim 0.782$ attained at $\phi\sim 0.58$, hence the largest relative error of the edge height is about $4\%$.

\begin{figure}[htbp]
    \centering
    \includegraphics[width=10cm]{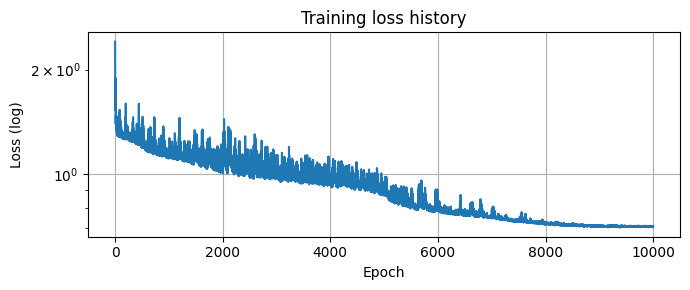}
    \caption{Loss function history for the standard minimal surface spanned by the light-like loop intersecting a Neumann boundary at $r=0.75$.}
    \label{fig:loss_square-N}
\end{figure}

Figure~\ref{fig:loss_square-N} shows the entire loss history, and table~\ref{tab:loss_square-N} shows the values of the loss components at the end of the training ($10^4$ epochs).
From initialization to convergence, the total loss decreased from $2.41$ to $7.09\times10^{-1}$ ($\sim70.6\%$ reduction).
$\mathrm{PDE}(R)$
dropped by $\sim93.7\%$ ($1.44\!\to\!9.10\times10^{-2}$), whereas
$\mathrm{PDE}(y_0)$ declined by $\sim24.6\%$ ($7.44\times10^{-1}\!\to\!5.61\times10^{-1}$).
At the end of the training, 
the interior residual $\mathrm{PDE}(y_0)$ dominates at convergence ($\approx79\%$ of the total),
with $\mathrm{PDE}(R)$ contributing $\approx 13\%$, including the numerical coefficient introduced in the loss function formula~\eqref{loss_square-N}. Among the boundary terms,
the largest share is the slope condition at $\theta_0$, $\partial_\theta(R_\theta\sin)$ ($\approx2.9\%$), and the others are relatively small ($\lesssim 1\%$).

\begin{table}[htbp]
  \centering
  \setlength{\tabcolsep}{7pt}
  \renewcommand{\arraystretch}{1.12}
  \begin{tabular}{lcc}
    \hline
    \textbf{Loss component} & \textbf{Value} & \textbf{Share of Total} \\
    \hline
    Total                                           & $7.09\times 10^{-1}$ & $100.00\%$ \\
    PDE($y_0$)                                            & $5.61\times 10^{-1}$ & $79.20\%$ \\
    PDE($R$) $(\times 5)$                                           & $9.10\times 10^{-2}$ & $12.85\%$ \\
    $\theta_0$-constraint $(R\cos\theta=r_0)$ $(\times 20)$       & $8.02\times 10^{-3}$ & $1.13\%$ \\
    BC @ $\theta=\theta_0$: $\partial_\theta(R\sin\theta)$
                                                    & $2.04\times 10^{-2}$ & $2.88\%$ \\
    BC @ $\theta=\theta_0$: $\partial_\theta y_0$        & $3.28\times 10^{-3}$ & $0.46\%$ \\
    BC $f_r$ [$\phi=0$]                             & $9.60\times 10^{-3}$ & $1.36\%$ \\
    BC $f_r$ [$\phi=\pi/4$]                         & $2.59\times 10^{-3}$ & $0.37\%$ \\
    BC $\tilde f_{y_0}$ [$\phi=0$]                  & $1.08\times 10^{-4}$ & $0.02\%$ \\
    BC $\tilde f_{y_0}$ [$\phi=\pi/4$]              & $1.23\times 10^{-2}$ & $1.74\%$ \\
    \hline
  \end{tabular}
  \caption{The values of the loss components in~\eqref{loss_square-N} at the end of the training ($10^4$ epochs). The values include the numerical coefficient included in the loss function formula~\eqref{loss_square-N}.}
  \label{tab:loss_square-N}
\end{table}

In summary, the network simultaneously solves the interior PDEs and the moving-boundary geometry, achieving smooth fields from the AdS boundary at $\theta=\pi/2$ up to the moving boundary at $\theta=\theta_0(\phi)$.
As is the case for the standard minimal surface, we observed that the PDE losses, particularly that for $y_0$, are the most significant obstacles against realizing a numerical solution with reasonable accuracy. 
For further study, improvements of the numerical scheme and the analytical treatment behind it would be the next step, which we defer to future works.

\section{Summary and discussions}
\label{sec:summary}

In this work, we have developed and demonstrated a physics-informed neural network (PINN) framework for solving boundary value problems involving minimal surfaces in curved geometry, with a particular focus on the challenges posed by singularities and moving boundaries in ordinary and partial differential equations. While our primary motivation stems from the context of the AdS/CFT correspondence in theoretical physics, the methods and insights presented here are widely applicable to boundary value problems encountered across mathematics, engineering, and other fields of the natural sciences.

A central theme of our study is the effective treatment of singularities—such as those present at the AdS boundary or at coordinate axes—which are well-known to impede conventional numerical approaches. 
The reason why we consider such cases is two-fold: first, these problems are popular settings in gravity and high energy physics research, and second, to test the novel numerical technology, those difficult but well-defined physics problems are most suitable.
By carefully designing the neural network ansatz with appropriate baseline and envelope functions, and by incorporating singular behavior directly into the formulation, we achieved robust and accurate PINN solutions even in the vicinity of such singular points. Additionally, we addressed the challenge of moving boundaries, which naturally arise in problems where part of the boundary is not fixed but determined dynamically by physical or geometric constraints. 

Our framework supports both “soft” (loss-based) and “hard” (formulation-based) imposition of boundary conditions, including those associated with moving boundaries. We found that the hard enforcement approach, in particular, enables the automatic satisfaction of complicated or dynamically determined boundary conditions, often resulting in faster convergence and improved numerical stability.
We also introduced a method to impose the boundary condition just by switching the ODE/PDE loss term into that for the boundary condition using the tophat window function. This method is easily implemented just by rewriting the code for a problem without the boundary, while it typically requires longer training for the PINN calculation to converge.

Despite these advances, several subtleties require further consideration. The design of the network ansatz and the implementation of boundary conditions become increasingly complex as the dimensionality of the problem or the strength of the singularities at the boundary increases.
While hard enforcement improves stability,
it limits the generality of the ansatz for more complex geometries.
Furthermore, as highlighted by our comparison of different enforcement strategies, minor discrepancies can persist near boundaries or singular points, indicating the need for further refinements in regularization and network architecture.

Looking ahead, the approaches developed here open several promising directions for future research.
One is to apply and adapt these methods to a broader array of boundary value problems, including those governed by nonlinear ODEs and PDEs in various research fields where singularities and moving boundaries frequently arise (e.g., in interface evolution, reaction-diffusion systems, membrane mechanics, or dynamics of astrophysical bodies).
Another important direction is the development of more flexible and adaptive PINN architectures, possibly by improving the solution ansatz and also introducing more advanced techniques such as domain decomposition or adaptive sampling, to efficiently handle stronger singularities or complex boundary geometries. 
Further work could also explore hybrid methods that combine the strengths of soft and hard boundary enforcement or integrate PINN-based approaches with established numerical techniques.
Finally, extending this framework to address initial value problems and time-dependent boundary value problems would significantly broaden its applicability in both fundamental and applied sciences.


\section*{Acknowledgments}
The work of K.~H.~was supported in part by JSPS KAKENHI Grant No.~JP22H01217, JP22H05111 and JP22H05115.
The work of N.~T.~was supported in part by JSPS KAKENHI Grant No.~JP21H05189, JP22H05111 and 25K07282.


\appendix

\section{Construction of solution ansatz for $R$ and $y_0$}
\label{App:ansatz}

The construction method is as follows. First, construct series solutions near the corner, introducing a coordinate system that regularizes the solutions. Next, based on the form of the series solutions, we define the baseline and envelope functions in the regularizing coordinates. Then, we obtain their expressions in the $(\theta,\phi)$ coordinates by means of a coordinate transformation. 

\subsection{Regularizing coordinates and series solutions}

We construct the series solutions of the Euler-Lagrange equations \eqref{EoM_y_surface}, \eqref{EoM_r_surface} as the first step for constructing the baseline and envelope functions.
The series solutions are most easily obtained in the Cartesian coordinates $(y_1,y_2)$. A subtlety of these coordinates is that the solution is expressed in terms of the square root of the coordinates, as we can observe in the exact solution \eqref{exact-soln}. To circumvent this issue, we introduce coordinates $Y_1, Y_2$ that regularize the solutions near the edge as
\begin{equation}
    y_i = \sin Y_i\qquad (i=1,2)\,.
\end{equation}
In terms of the new coordinates $Y_i$, the series solutions near the edge are expressed as polynomials of them. 
As an illustration, let us expand the exact solution \eqref{exact-soln} near one of the corner $(y_1,y_2) = (-1, -1)$.
Defining  displacements $\Delta Y_i$ by
\begin{equation}
    Y_i = -\frac{\pi}{2} + \Delta Y_i\,,
\end{equation}
the exact solution is expanded for small $\Delta Y_i$ as
\begin{align}
    y_0 &= \sin Y_1 \sin Y_2 
    =
    1 - \frac{1}{2}\left( \Delta Y_1^2 + \Delta Y_2^2 \right)
    +\frac{1}{24}\left(\Delta Y_1^4 + 6\Delta Y_1^2 \Delta Y_2^2 + \Delta Y_2^2\right) 
    + \cdots,
    \\
    r &= \cos Y_1 \cos Y_2 
    = 
    \Delta Y_1 \Delta Y_2 - \frac{1}{6} \Delta Y_1 \Delta Y_2 \left( \Delta Y_1^2 + \Delta Y_2^2 \right) + \cdots.
\end{align}

We construct the series solution around the corner $(y_1,y_2) = (-1, -1)$ using the variables introduced above. Expressing the general solutions as 
\begin{equation}
y_0 = \sum_{i,j=0}^\infty C^{y_0}_{i,j} \Delta Y_1{}^i\Delta Y_2{}^j\,,
r = \sum_{i,j=0}^\infty C^{r}_{i,j} \Delta Y_1{}^i\Delta Y_2{}^j \,,
\end{equation}
where $C^{y_0}_{i,j}$, $C^{r}_{i,j}$ are constant coefficients.
Plugging them into the equations \eqref{EoM_y_surface} and \eqref{EoM_r_surface},
expanding them for small $\Delta Y_i$, and imposing the boundary conditions given in section~\ref{sec:BC_square}, we find the general series solutions are given by
\begin{align}
    y_0 &= 
    1 - \frac{1}{2}\left( \Delta Y_1^2 + \Delta Y_2^2 \right)
    +\frac{1}{24}\left(\Delta Y_1^4 + \Delta Y_2^2\right) 
    + C^{y_0}_{2,2} \Delta Y_1{}^2 \Delta Y_2{}^2 
    + \cdots,
    \label{series-y0_square}
    \\
    r &= 
    \Delta Y_1 \Delta Y_2 
    + C^{r}_{1,3} \Delta Y_1 \Delta Y_2 \left(\Delta Y_1{}^2 + \Delta Y_2{}^2\right)+ \cdots,
    \label{series-r_square}
\end{align}
where $C^{y_0}_{2,2}$ and $C^{r}_{1,3}$ remain undetermined by the boundary conditions in section~\ref{sec:BC_square}.
In \eqref{series-r_square}, we have set $C^{r}_{3,1} = C^{r}_{1,3}$ based on the symmetry of the problem setting.
The coefficients $C^{r,y_0}_{i,j}$ at higher order are uniquely determined in terms of $C^{y_0}_{2,2}$ and $C^{r}_{1,3}$.

\subsection{Solution ansatz}

The form of the series solution \eqref{series-y0_square}, \eqref{series-r_square} suggests that we may express the general solution as, at least for small $\Delta Y_i$,
\begin{align}
    y_0(\Delta Y_1, \Delta Y_2) &=
    y_0^{\text{exact}} + \Delta Y_1{}^2\Delta Y_2{}^2 \, f_{y_0}(\Delta Y_1, \Delta Y_2)\,, \\
    r(\Delta Y_1, \Delta Y_2) &=
    r^{\text{exact}} + \Delta Y_1 \Delta Y_2 \left( \Delta Y_1^2 + \Delta Y_2^2 \right)f_r(\Delta Y_1, \Delta Y_2)\,,    
\end{align}
where $y_0^{\text{exact}}, r^{\text{exact}}$ are the exact solution \eqref{exact-soln} and $f^{(y_0)}(\Delta Y_1, \Delta Y_2), f_r(\Delta Y_1, \Delta Y_2)$ are the free functions that become $O(1)$ for small $\Delta Y_i$. 
By expanding this ansatz for small $\Delta Y_i$, we recover the series expansion \eqref{series-y0_square} and \eqref{series-r_square}, where $C^{y_0}_{2,2}$ and $C^{r}_{1,3}$ are given by $f^{(y_0)}(0,0)$ and $f_r(0,0)$ shifted by the series coefficients obtained from the exact solution $y_0^{\text{exact}}$ and $r^{\text{exact}}$.

We generalize the above definition so that it works even when $\Delta Y_i$ are not necessarily small by replacing $\Delta Y_i$ in the coefficients with $\sin \Delta Y_i = - \cos Y_i$. It results in
\begin{align}
    y_0(Y_1, Y_2) &=
    y_0^{\text{exact}} + \cos^2 Y_1 \cos^2 Y_2 \, f_{y_0}(Y_1, Y_2)\,, \\
    r(Y_1, Y_2) &=
    r^{\text{exact}} + \cos Y_1 \cos Y_2 \left( \cos^2 Y_1 + \cos^2 Y_2 \right)f_r(Y_1, Y_2)\,.
\end{align}
In terms of the original coordinates $y_i$, they are expressed as
\begin{align}
    y_0(y_1,y_2) &=
    y_1 y_2 + (1-y_1{}^2) (1-y_2{}^2) \, f_{y_0}(y_1, y_2)\,,
    \label{ansatz_y0}
    \\
    r(y_1, y_2) &=
    \sqrt{(1-y_1{}^2)(1-y_2{}^2)}
    + \sqrt{(1-y_1{}^2)(1-y_2{}^2)} \left(2- y_1{}^2 - y_2{}^2 \right) f_r(y_1, y_2)
    \,.
    \label{ansatz_r}
\end{align}
Here, we assume that $f_{y_0}, f_r$ are smooth functions that becomes $O(1)$ near the boundary $C$.

\subsection{Conversion to the spherical coordinates}

Next task is to convert the solution ansatz \eqref{ansatz_y0}, \eqref{ansatz_r} into the spherical coordinates \eqref{sph-coords_R}.
For this purpose, we plug in $y_1, y_2$ expressed in the spherical coordinates into the ansatz \eqref{ansatz_y0}, \eqref{ansatz_r},
which yields
\begin{align}
    y_0 &=
    y_1 y_2 + (1-y_1{}^2) (1-y_2{}^2) \, f_{y_0}(y_1, y_2)
    \Bigr|_{\substack{y_1 = R \sin\theta\cos\phi\\y_2 = R \sin\theta\sin\phi}}
    \,,
    \label{ansatz_y0-2}
    \\
    r &= R \cos\theta
    \notag \\
    &=
    \sqrt{(1-y_1{}^2)(1-y_2{}^2)}
    + \sqrt{(1-y_1{}^2)(1-y_2{}^2)} \left(2- y_1{}^2 - y_2{}^2 \right) f_r(y_1, y_2)
    \Bigr|_{\substack{y_1 = R \sin\theta\cos\phi\\y_2 = R \sin\theta\sin\phi}}
    \,.
    \label{ansatz_r-2}
\end{align}
Then, we assume that $y_0(\theta,\phi)$ and $R(\theta,\phi)$ are expressed as
\begin{align}
    y_0(\theta,\phi) &= \mathcal{B}_{y_0}(\theta,\phi) + \Delta f_{y_0}(\theta,\phi)\,,
    \label{ansatz_y0-Deltaf}
    \\
    R(\theta,\phi) &= \mathcal{B}_{R}(\theta,\phi) + \Delta f_R(\theta,\phi)\,,
    \label{ansatz_R-Deltaf}
\end{align}
where $\mathcal{B}_{y_0}, \mathcal{B}^{(r)}$ are the baseline functions and $\Delta f_{y_0}, \Delta f_R$ are the differences between the baseline function and $y_0, R$.

We first plug \eqref{ansatz_R-Deltaf} into \eqref{ansatz_r-2}, and it yields a nonlinear equation involving $R, \tilde f, \Delta f$.
By solving this equation under the assumption that $\tilde f$ and $\Delta f$ are small quantities of the same order, we find
\begin{equation}
    R(\theta,\phi) = \mathcal{B}_{R}(\theta,\phi) + \mathcal{E}_R(\theta,\phi) \times  f_r(\theta,\phi)\, ,    \label{ansatz-R_square}
\end{equation}
where
\begin{equation}
    \mathcal{B}_{R} \equiv \sqrt{\frac{2}{1 +\rho}}\,,
    \qquad
    \mathcal{E}_R \equiv
    \frac{2\sqrt{2}\cos^2\theta\left(1 + \frac{\cos^2\theta}{\rho}\right) }{(1 + \rho)^{3/2}}
    \,,
\end{equation}
for which we introduced
\begin{equation}
    \rho \equiv \sqrt{1 - \sin^4 \theta \sin^2(2\phi)}\,.
\end{equation}

Plugging \eqref{ansatz_y0-Deltaf} and \eqref{ansatz-R_square} into \eqref{ansatz_y0-2}, and solving it assuming that $f_r, f_{y_0}, \Delta f_{y_0}$ are small quantities at the same order, we obtain
\begin{equation}
    y_0(\theta,\phi) =
    \mathcal{B}_{y_0}(\theta,\phi)
    + \mathcal{E}_{y_0; y_0}(\theta,\phi) f_{y_0}(\theta,\phi) 
    + \mathcal{E}_{y_0; r}(\theta,\phi) f_{r}(\theta,\phi) \,,
    \label{ansatz-y0_square}
\end{equation}
where
\begin{equation}
    \mathcal{B}_{y_0} \equiv \frac{\sin^2\theta \sin(2\pi)}{1 + \rho}\,,
    \quad
    \mathcal{E}_{y_0; y_0} \equiv \frac{2 \cos^2 \theta}{1+\rho}\,,
    \quad
    \mathcal{E}_{y_0; r} \equiv \frac{2 \sin^2\theta \sin(2\phi)}{\sqrt{2+2\rho}} \times \mathcal{E}_R(\theta,\phi)
    \,.
\end{equation}

In our numerical code, we use \eqref{ansatz-R_square} and \eqref{ansatz-y0_square} to express $R$ and $y_0$, and assume that $f_R, f_{y_0}$ are smooth and finite everywhere.


\section{Another PINN approach to minimal surface with light-like boundary}
\label{app:square-alt}

In this appendix, we present another attempt to apply PINN to find the minimal surface treated in section \ref{sec:light-like_loop}.
The message of this appendix is that the implementation of PINN for a minimal surface problem actually allows a variety of choices for the structure of the loss functions and treatment of the boundary conditions and singularities. As we see below, even a careful treatment of the PINN architecture often fails to reproduce the exact solution. The strategy delivered in the main text of this paper has been discovered by multiple of try and errors based on empirical findings in various applications of PINN. The readers are encouraged to explore various approaches to pile up effective methods to pave the road for the new world of the novel PINN technology.

\subsection{Exact solution and boundary conditions}
The exact solution for the minimal surface with the light-like boundary  was given in \eqref{exact-soln}.
%
The function $y_0$ describes the embedding of the minimal surface in AdS space, while $r$ represents the radial coordinate in terms of the boundary coordinates $(y_1, y_2)$.
\begin{figure}[htbp]
        \centering
        \subfigure[$r(y_1, y_2) = \sqrt{(1-y_1^2)(1-y_2^2)}$]{\includegraphics[width=7cm]{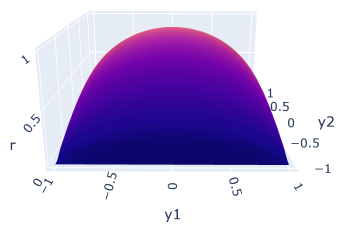}}
        \subfigure[$y_0(y_1, y_2) = y_1y_2$]{\includegraphics[width=7cm]{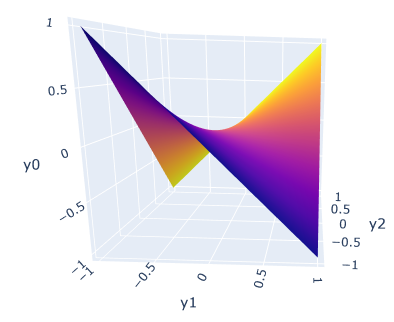}}
        \caption{Exact solution surfaces for the 4-point function. (a) The radial coordinate $r$ exhibiting characteristic behavior near the boundaries $y_1, y_2 = \pm 1$. (b) The embedding function $y_0$ showing a saddle-like structure. These exact solutions serve as the target for our PINN validation.}
\label{fig:exact_solution}
    \end{figure}
This particular form emerges naturally from the Euler-Lagrange equations for minimal surfaces in AdS space with appropriate boundary conditions corresponding to light-like Wilson loops \eqref{eq:bc_rhombus}.
We again use the spherical coordinates defined in \eqref{sph-coords_R}.

As shown in figure~\ref{fig:exact_solution}, the exact solution exhibits complex geometric features including a saddle-like structure in the embedding function $Y_0$ and characteristic behavior near the boundaries. 
The AdS boundary exhibits a coordinate singularity, and the exact solution at the boundary $R(\theta=\pi/2,\phi)$ becomes discontinuous at the vertices of the square, which are located at $\phi=\pi/4, 3\pi/4, 5\pi/4, 7\pi/4$.
Due to this complexity, we avoid direct implementation at the exact boundary. Instead, we impose boundary conditions at a slightly interior point with $\theta_{\text{max}} = \frac{\pi}{2} \times 0.9$, which helps circumvent numerical instabilities while maintaining physical accuracy.

The boundary conditions for the radial coordinate $R$ and the embedding function $Y_0$ are implemented as follows:
\begin{align}
R(\theta=0,\Phi) &= R_{\text{tip}}, \\
\partial_\theta R(\theta=0,\phi) &= 0, \\
R(\theta_{\text{max}},\phi) &= R_{\text{sol}}(\theta_{\rm max},\phi) \,, \quad \\
\partial_\theta R(\theta_{\text{max}}, \phi) 
&= \partial_\theta R_{\text{sol}}(\theta_{\text{max}}, \phi)  \quad \\
Y_0(\theta=0,\Phi) &= Y_{0,\text{tip}}, \\
\partial_\theta Y_0(\theta=0,\Phi) &= 0, \\
Y_0(\theta_{\text{max}},\phi) &= Y_{0,\text{sol}}(\theta_{\text{max}}, \phi) \quad \, \\
\partial_\theta Y_0(\theta_{\text{max}}, \phi) 
&= \partial_\theta Y_{0,\text{sol}}(\theta_{\text{max}}, \phi)  \quad \,.
\end{align}
Here $R_{\rm sol}$ and $Y_{0,\rm {sol}}$ represent the exact solutions obtained by the coordinate transformation from 
\eqref{exact-soln} and $R_{\rm tip}$ is introduced as \eqref{eq:Rtip} as well as $Y_{0,\rm{tip}}$.

To accurately capture the minimal surface with light-like boundary while ensuring smooth boundary transitions, we employ a \emph{hard enforcing} approach similar to that described in section \ref{sec:light-like_loop}.
This method directly incorporates boundary conditions into the functional form of our solution, rather than enforcing them through additional loss terms.

Following this hard enforcing strategy, both the radial coordinate $R$ and the embedding function $Y_0$ are parameterized as:
\begin{align}
R(\theta,\Phi) &= R_{\text{tip}} + \frac{\theta^2}{\theta^2_{\text{max}}} \left(3R_{\text{sol}}(\theta_{\text{max}}, \phi)  - 3R_{\text{tip}} - \theta_{\text{max}}\partial_\theta R_{\text{sol}}(\theta_{\text{max}}, \phi) \right) \notag\\
&\quad + \frac{\theta^3}{\theta^3_{\text{max}}} \left(2R_{\text{tip}} - 2R_{\text{sol}}(\theta_{\text{max}}, \phi)  + \theta_{\text{max}}\partial_\theta R_{\text{max}}\right) 
+ \mathcal{E}(\theta) \times R_{\text{NN}}(\theta,\Phi), \\
Y_0(\theta,\Phi) &= Y_{0,\text{tip}} + \frac{\theta^2}{\theta^2_{\text{max}}} \left(3Y_{0,\text{sol}}(\theta_{\text{max}}, \phi)  - 3Y_{0,\text{tip}} - \theta_{\text{max}}\partial_\theta Y_{0,\text{sol}}(\theta_{\text{max}}, \phi) \right) \notag\\
&\quad 
+ \frac{\theta^3}{\theta^3_{\text{max}}} \left(2Y_{0,\text{tip}} - 2Y_{0,\text{sol}}(\theta_{\text{max}}, \phi)  + \theta_{\text{max}}\partial_\theta Y_{0,\text{sol}}(\theta_{\text{max}}, \phi) \right)
+ \mathcal{E}(\theta) \times Y_{0,\text{NN}}(\theta,\Phi),
\end{align}
where $R_{\text{NN}}$ and $Y_{0,\text{NN}}$ represent the neural network outputs that capture the deviations from the cubic polynomial baseline. 
The cubic baseline functions are constructed to automatically satisfy the boundary conditions at $\theta = 0$ and $\theta = \theta_{\text{max}}$.
On the other hand, by defining the envelope function $\mathcal{E}$ as
\begin{equation}
\mathcal{E}(\theta) = \theta^2 (\theta_{\text{max}} - \theta)^2\,,
\end{equation}
the envelope-modulated neural network terms provide the flexibility to capture the complex geometric features of the minimal surface in the interior domain.


Despite this careful implementation strategy, as will be demonstrated in section \ref{appendixB3}, the PINN approach still faces significant challenges in accurately reproducing the geometric features characteristic of these minimal surfaces with light-like boundary.

\subsection{Loss function for neural network training}

The total loss function used for training the neural network consists of three main components:

\begin{equation}
\mathcal{L}_{\text{total}} = \mathcal{L}_{\text{EOM}} + \lambda_{\text{BC}} \mathcal{L}_{\text{BC}} + \lambda_{\text{NG}} \mathcal{L}_{\text{NG}}
\end{equation}
where $\lambda_{\text{BC}} = 100$ and $\lambda_{\text{NG}} = 1000$ are weighting factors chosen to ensure proper enforcement of the boundary conditions and physical constraints, respectively. Each component is defined as follows:

\subsubsection{Equations of motion}
The first component enforces the equations of motion derived from the Nambu-Goto action \eqref{L_GC}:
\begin{equation}
\mathcal{L}_{\text{EOM}} = \sum_{\theta,\phi} |\text{PDE loss}|^2
\end{equation}
where the PDE loss represents the residual of the equations of motion evaluated at the sampling points. 
We have rescaled the equations of motion by an overall factor of $G^{3/2}$ to ensure numerical stability and proper normalization, where $G$ represents the determinant of the induced metric appearing under the square root in the Nambu-Goto action.
This term ensures that the neural network solution satisfies the fundamental dynamical equations governing the minimal surface.

\subsubsection{Periodic boundary conditions}
The second component, weighted by $\lambda_{\text{BC}} = 100$, ensures the periodic boundary conditions in the $\phi$ direction:
\begin{equation}
\begin{aligned}
\mathcal{L}_{\text{BC}} = &\sum_{\theta} \left[ |R(\theta,\phi=0) - R(\theta,\phi=2\pi)|^2 + |R_t(\theta,\phi=0) - R_t(\theta,\phi=2\pi)|^2 \right. \\
&\left. + |Y_0(\theta,\phi=0) - Y_0(\theta,\phi=2\pi)|^2 + |Y_{0,t}(\theta,\phi=0) - Y_{0,t}(\theta,\phi=2\pi)|^2 \right]
\end{aligned}
\end{equation}
These terms enforce the periodicity of the embedding functions $R(\theta,\phi)$ and $Y_0(\theta,\phi)$, as well as their time derivatives $R_t(\theta,\phi)$ and $Y_{0,t}(\theta,\phi)$. The relatively large weighting factor $\lambda_{\text{BC}} = 100$ reflects the importance of maintaining exact periodicity for the physical validity of the solution.

\subsubsection{Nambu-Goto action constraint}
The third component, weighted by $\lambda_{\text{NG}} = 1000$, ensures the physical validity of the solution by requiring that the argument inside the square root of the Nambu-Goto action remains non-negative:
\begin{equation}
\mathcal{L}_{\text{NG}} = \sum_{\theta,\phi} [\text{ReLU}(-G)]^2
\end{equation}
where $G$ is the determinant appearing under the square root in the Nambu-Goto action\eqref{L_GC}, and ReLU is the rectified linear unit function defined as:
\begin{equation}
\text{ReLU}(x) = \max(0, x)
\end{equation}

This penalty term becomes active only when $G < 0$, preventing unphysical solutions where the induced metric would have an imaginary area element. The large weighting factor $\lambda_{\text{NG}} = 1000$ strongly penalizes any violations of this fundamental physical constraint, ensuring that the neural network converges to a physically meaningful solution.
The choice of these specific weighting factors was determined through extensive numerical experiments, balancing the need to satisfy all constraints while maintaining stable convergence during the training process. 

\begin{figure}[htbp]
    \centering
    \subfigure[$r(y_1,y_2)$]{\includegraphics[width=7cm]{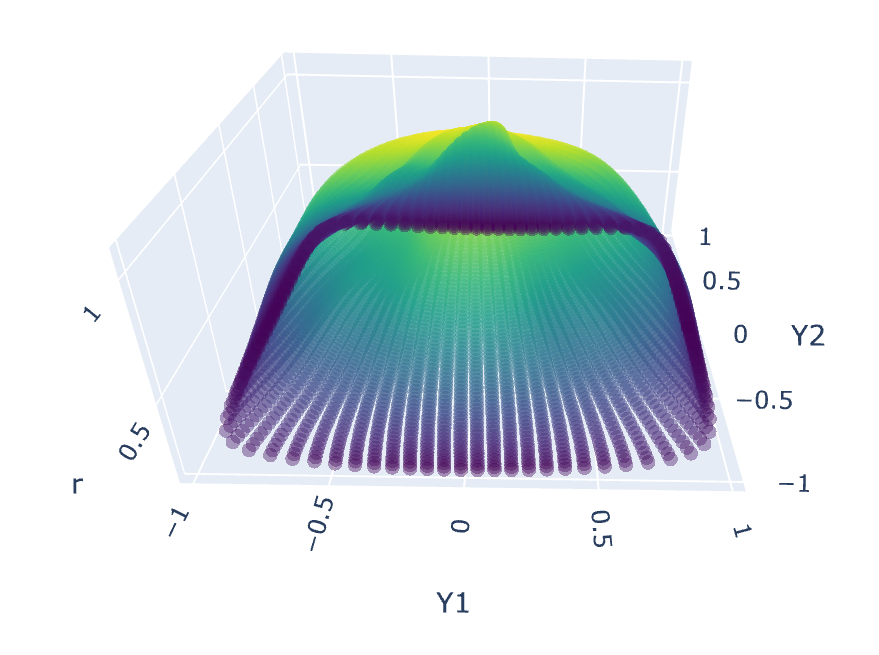}}
    \subfigure[$r(y)$ at $\phi = 0$]{\includegraphics[width=7cm]{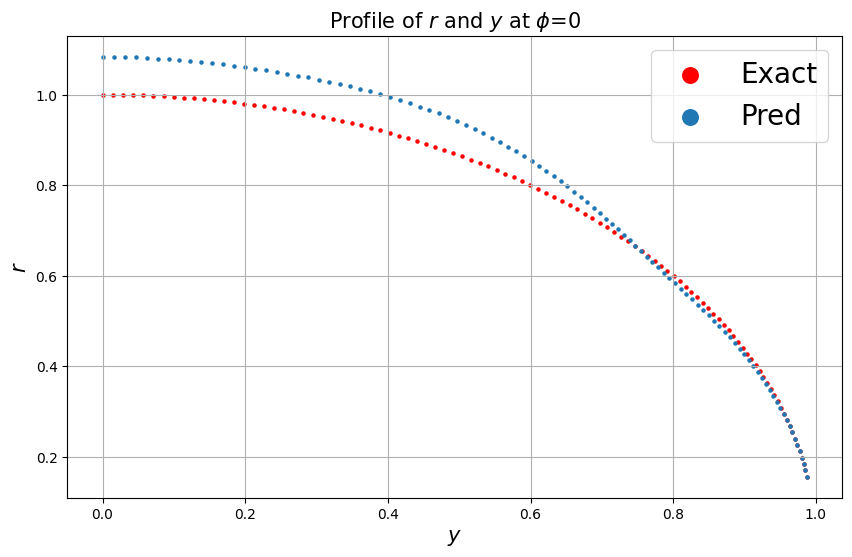}}
    \subfigure[$r(y)$ at $\phi = \pi/4$]{\includegraphics[width=7cm]{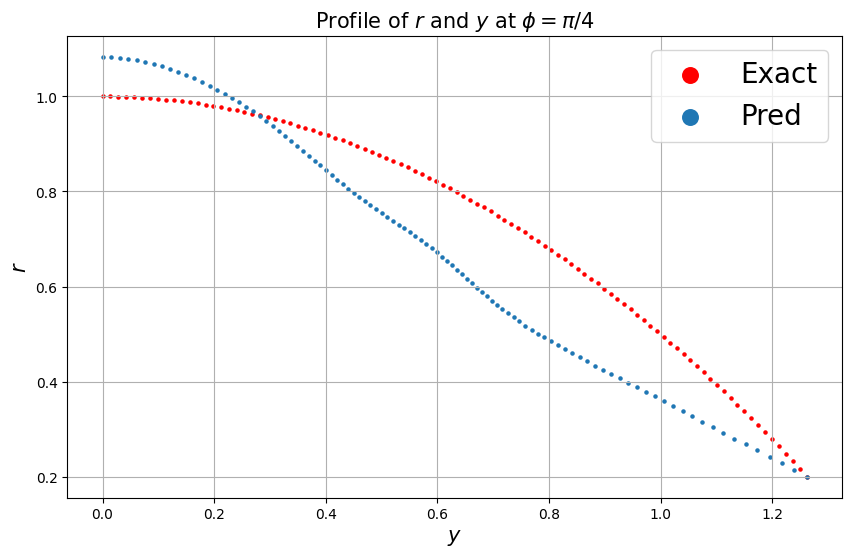}}
    \subfigure[$Y_0(y_1,y_2)$]{\includegraphics[width=7cm]{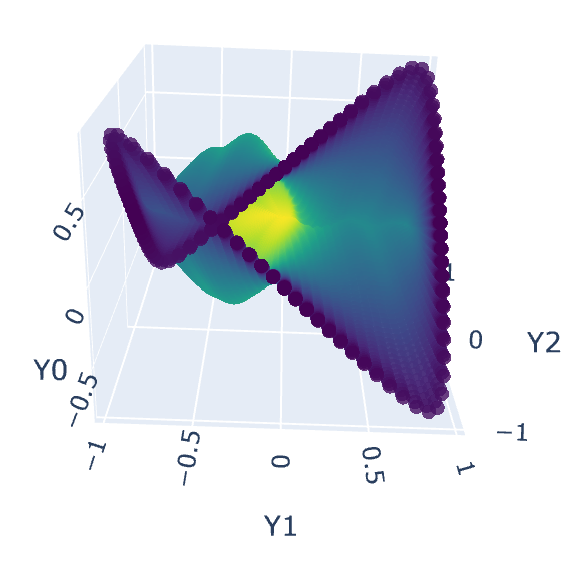}}
    \caption{PINN results for the radial coordinate $r(y_1,y_2)$ and the function $Y_0(y_1,y_2)$ in the calculation of minimal surface with light-like boundary. (a) $(y_1,y_2)$, showing deviations at corner positions. (b) $r(y)$ at $\phi = 0$, where $y=\sqrt{y_1^2+y_2^2}$ and the exact solution is drawn by red curves. The solution remains relatively accurate. (c) $r(y)$ at $\phi = \pi/4$, exhibiting significant errors near the square corners. (d) $Y_0(y_1,y_2$) function with visible distortions. }
    \label{fig:r_fig}
\end{figure}

\subsection{Results and profile characteristics}
\label{appendixB3}

As shown in figure~\ref{fig:r_fig}, our PINN implementation reveals significant challenges in accurately reproducing the minimal surfaces with light-like boundary, particularly at critical geometric features:
\begin{description}
\item[Profile of $r(y_1,y_2)$:] The radial profile shows substantial deviations from the exact solution at the four corners of the square ($\phi = \pi/4, 3\pi/4, 5\pi/4, 7\pi/4$). While the PINN captures the general shape with characteristic concavities at these positions, the accuracy deteriorates significantly near these singular points, indicating the network's difficulty in handling sharp geometric transitions.

\item[Profile of $Y_0(y_1,y_2)$:]
The $Y_0$ profile exhibits similar distortions at the four corners of the square ($\phi = \pi/4, 3\pi/4, 5\pi/4, 7\pi/4$), where the network struggles to maintain the proper saddle-like structure. These regions correspond to the corners of the boundary, suggesting that the standard PINN approach has fundamental limitations when dealing with non-smooth boundary geometries.
\end{description}

These results demonstrate the limitations of standard PINN approaches when dealing with minimal surfaces that exhibit sharp variations or singular behavior at the boundaries. The difficulties encountered in this light-like boundary case motivate the development of more sophisticated boundary treatment methods and the alternative approaches explored in the main text of this paper.

\bibliographystyle{JHEP}
\bibliography{paper2}

\providecommand{\href}[2]{#2}\begingroup\raggedright\begin{thebibliography}{10}

\bibitem{Maldacena:1997re}
J.M.~Maldacena, \emph{{The Large N limit of superconformal field theories and supergravity}}, \href{https://doi.org/10.4310/ATMP.1998.v2.n2.a1}{\emph{Adv. Theor. Math. Phys.} {\bfseries 2} (1998) 231} [\href{https://arxiv.org/abs/hep-th/9711200}{{\ttfamily hep-th/9711200}}].

\bibitem{RAISSI2019686}
M.~Raissi, P.~Perdikaris and G.~Karniadakis, \emph{Physics-informed neural networks: A deep learning framework for solving forward and inverse problems involving nonlinear partial differential equations}, \href{https://doi.org/https://doi.org/10.1016/j.jcp.2018.10.045}{\emph{Journal of Computational Physics} {\bfseries 378} (2019) 686}.

\bibitem{NASCIMENTO2020103996}
R.G.~Nascimento, K.~Fricke and F.A.~Viana, \emph{A tutorial on solving ordinary differential equations using python and hybrid physics-informed neural network}, \href{https://doi.org/https://doi.org/10.1016/j.engappai.2020.103996}{\emph{Engineering Applications of Artificial Intelligence} {\bfseries 96} (2020) 103996}.

\bibitem{zhou2023approximatinghighdimensionalminimalsurfaces}
S.~Zhou and X.~Ye, \emph{Approximating high-dimensional minimal surfaces with physics-informed neural networks},  2023.

\bibitem{Kabasi2023PINN}
S.~Kabasi, A.L.~Marbaniang and S.~Ghosh, \emph{Physics-informed neural networks for the form-finding of tensile membranes by solving the euler^^e2^^80^^93lagrange equation of minimal surfaces}, \href{https://doi.org/10.1016/j.tws.2022.110309}{\emph{Thin-Walled Structures} {\bfseries 182} (2023) 110309}.

\bibitem{peng2021idrlnetphysicsinformedneuralnetwork}
W.~Peng, J.~Zhang, W.~Zhou, X.~Zhao, W.~Yao and X.~Chen, \emph{Idrlnet: A physics-informed neural network library},  2021.

\bibitem{Mishra2025TPMS}
A.~Mishra, \emph{Machine learning-driven optimization of tpms architected materials using simulated annealing}, \href{https://doi.org/10.1007/s44379-024-00001-z}{\emph{Machine Learning for Computational Science and Engineering} {\bfseries 1} (2025) 1}.

\bibitem{Maldacena:1998im}
J.M.~Maldacena, \emph{{Wilson loops in large N field theories}}, \href{https://doi.org/10.1103/PhysRevLett.80.4859}{\emph{Phys. Rev. Lett.} {\bfseries 80} (1998) 4859} [\href{https://arxiv.org/abs/hep-th/9803002}{{\ttfamily hep-th/9803002}}].

\bibitem{Rey:1998bq}
S.-J.~Rey, S.~Theisen and J.-T.~Yee, \emph{{Wilson-Polyakov loop at finite temperature in large N gauge theory and anti-de Sitter supergravity}}, \href{https://doi.org/10.1016/S0550-3213(98)00471-4}{\emph{Nucl. Phys. B} {\bfseries 527} (1998) 171} [\href{https://arxiv.org/abs/hep-th/9803135}{{\ttfamily hep-th/9803135}}].

\bibitem{Alday:2007hr}
L.F.~Alday and J.M.~Maldacena, \emph{{Gluon scattering amplitudes at strong coupling}}, \href{https://doi.org/10.1088/1126-6708/2007/06/064}{\emph{JHEP} {\bfseries 06} (2007) 064} [\href{https://arxiv.org/abs/0705.0303}{{\ttfamily 0705.0303}}].

\bibitem{Hashimoto2025instanton}
K.~Hashimoto, K.~Kyo, M.~Murata, G.~Ogiwara and N.~Tanahashi, \emph{{Gluon scattering amplitudes with instantons and minimal surfaces with topology change}},  \href{https://arxiv.org/abs/2509.10865}{{\ttfamily 2509.10865}}.

\bibitem{doi:10.1061/(ASCE)EM.1943-7889.0001947}
C.~Rao, H.~Sun and Y.~Liu, \emph{Physics-informed deep learning for computational elastodynamics without labeled data}, \href{https://doi.org/10.1061/(ASCE)EM.1943-7889.0001947}{\emph{Journal of Engineering Mechanics} {\bfseries 147} (2021) 04021043}.

\bibitem{jin2020unsupervised}
H.~Jin, M.~Mattheakis and P.~Protopapas, \emph{Unsupervised neural networks for quantum eigenvalue problems}, {\emph{arXiv preprint arXiv:2010.05075} (2020) }.

\bibitem{PhysRevE.105.065305}
M.~Mattheakis, D.~Sondak, A.S.~Dogra and P.~Protopapas, \emph{Hamiltonian neural networks for solving equations of motion}, \href{https://doi.org/10.1103/PhysRevE.105.065305}{\emph{Phys. Rev. E} {\bfseries 105} (2022) 065305}.

\bibitem{Luna:2022rql}
R.~Luna, J.~Calder{\'o}n~Bustillo, J.J.S.~Mart{\'\i}nez, A.~Torres-Forn{\'e} and J.A.~Font, \emph{{Solving the Teukolsky equation with physics-informed neural networks}}, \href{https://doi.org/10.1103/PhysRevD.107.064025}{\emph{Phys. Rev. D} {\bfseries 107} (2023) 064025} [\href{https://arxiv.org/abs/2212.06103}{{\ttfamily 2212.06103}}].

\bibitem{Luna:2024spo}
R.~Luna, D.D.~Doneva, J.A.~Font, J.-H.~Lien and S.S.~Yazadjiev, \emph{{Quasinormal modes in modified gravity using physics-informed neural networks}}, \href{https://doi.org/10.1103/PhysRevD.109.124064}{\emph{Phys. Rev. D} {\bfseries 109} (2024) 124064} [\href{https://arxiv.org/abs/2404.11583}{{\ttfamily 2404.11583}}].

\bibitem{TSENG2023112359}
Y.-H.~Tseng, T.-S.~Lin, W.-F.~Hu and M.-C.~Lai, \emph{A cusp-capturing pinn for elliptic interface problems}, \href{https://doi.org/https://doi.org/10.1016/j.jcp.2023.112359}{\emph{Journal of Computational Physics} {\bfseries 491} (2023) 112359}.

\bibitem{hu2024solving}
T.~Hu, B.~Jin and Z.~Zhou, \emph{Solving poisson problems in polygonal domains with singularity enriched physics informed neural networks}, {\emph{SIAM Journal on Scientific Computing} {\bfseries 46} (2024) C369}.

\bibitem{Cayuso:2024jau}
R.~Cayuso, M.~Herrero-Valea and E.~Barausse, \emph{{Deep learning solutions to singular ordinary differential equations: From special functions to spherical accretion}}, \href{https://doi.org/10.1103/PhysRevD.111.064082}{\emph{Phys. Rev. D} {\bfseries 111} (2025) 064082} [\href{https://arxiv.org/abs/2409.20150}{{\ttfamily 2409.20150}}].

\bibitem{Dobashi:2008ia}
S.~Dobashi, K.~Ito and K.~Iwasaki, \emph{{A Numerical Study of Gluon Scattering Amplitudes in N=4 Super Yang-Mills Theory at Strong Coupling}}, \href{https://doi.org/10.1088/1126-6708/2008/07/088}{\emph{JHEP} {\bfseries 07} (2008) 088} [\href{https://arxiv.org/abs/0805.3594}{{\ttfamily 0805.3594}}].

\bibitem{Dobashi:2009sj}
S.~Dobashi and K.~Ito, \emph{{Discretized Minimal Surface and the BDS Conjecture in N=4 Super Yang-Mills Theory at Strong Coupling}}, \href{https://doi.org/10.1016/j.nuclphysb.2009.04.005}{\emph{Nucl. Phys. B} {\bfseries 819} (2009) 18} [\href{https://arxiv.org/abs/0901.3046}{{\ttfamily 0901.3046}}].

\end{thebibliography}\endgroup

%


\end{document}